\documentclass[prl,floatfix,twocolumn,superscriptaddress,showpacs,amsmath,amssymb,showpacs]{revtex4-1}

\usepackage{graphicx}
\usepackage{epstopdf}
\usepackage{epsfig}
\usepackage{epsf}
\usepackage{url}
\usepackage[USenglish]{babel}
\usepackage{hyperref}
\def\bcen{\begin{center}}
\def\ecen{\end{center}}
\allowdisplaybreaks
\usepackage{verbatim}
\usepackage{natbib}
\usepackage{amsmath, nccmath}
\usepackage[export]{adjustbox}
\usepackage{dcolumn}
\usepackage{bm}
\usepackage{bbm}
\usepackage{lipsum}

\def\bcen{\begin{center}}
\def\ecen{\end{center}}

\usepackage[section]{placeins}

\begin{document}
	
	\title{Nature of the photo-induced metallic state in monoclinic VO$_2$}
	
	\author{Jiyu Chen}
	\affiliation{Department of Physics, University of Fribourg, 1700 Fribourg, Switzerland}
	\author{Francesco Petocchi}
	\affiliation{Department of Physics, University of Geneva, 1211  Geneva, Switzerland}
	\author{Viktor Christiansson}
    \affiliation{Department of Physics, University of Fribourg, 1700 Fribourg, Switzerland}
	\author{Philipp Werner}
	%\email{philipp.werner@unifr.ch}
	\affiliation{Department of Physics, University of Fribourg, 1700 Fribourg, Switzerland}

\begin{abstract} 
The metal-insulator transition of VO$_2$, which in equilibrium is associated with a structural phase transition, has been intensively studied for decades. In particular, it is challenging to disentangle the role of Mott physics from dimerization effects in the insulating phase. Femtosecond time-resolved experiments showed that optical excitations can induce a transient metallic state in the dimerized phase, which is distinct from the known equilibrium phases. In this study, we combine non-equilibrium cluster dynamical mean-field theory with realistic first principles modeling to clarify the nature of this laser-induced metallic state. We show that the doublon-holon production by laser pulses with polarization along the V-V dimers and the subsequent inter-orbital reshuffling of the photo-carriers leads to a population of orbital-mixed states and the filling of the gap. The photo-induced metal state is qualitatively similar to a hot electronic state in the dimerized structure, and does not involve a collapse of the Mott gap.  
\end{abstract}

\maketitle
	
{\it Introduction.} Driving correlated electron materials out of their equilibrium state provides new perspectives on correlation phenomena and can shed light on competing or cooperative effects which are difficult to disentangle in equilibrium. Studies of the low-temperature phase of VO$_2$ provide an illustrative example of this general approach. VO$_2$ undergoes a metal-insulator transition (MIT) at $T_\text{MIT}= 340$~K, accompanied by a periodic lattice distortion \cite{morin1959}. While the metallic system above $T_\text{MIT}$ exhibits a rutile structure (R), below $T_\text{MIT}$ the dimerization of the V chain along the tetragonal $c$ axis and a lateral zigzag-type displacement result in a monoclinic insulator (M1). The V$^{4+}$ cations have a $3d^1$ configuration and are surrounded by oxygen octahedra, which splits the $3d$ levels into three low energy $t_{2g}$ orbitals and two high energy $e^{\sigma}_g$ orbitals. The tetragonal distortion further lifts the degeneracy of the $t_{2g}$ orbitals into an $a_{1g}$ orbital ($d_{x^2-y^2}$) and two $e_g^\pi$ orbitals ($d_{xz}$, $d_{yz}$)~\cite{eyert2002}. Above $T_\text{MIT}$, the $t_{2g}$ bands are partially filled, consistent with the metallic nature of the R phase. In the low-temperature M1 structure, the lattice distortion results in a bonding-antibonding splitting of the $a_{1g}$ bands and an upward-shift of the $e^{\pi}_g$ bands.  Goodenough \cite{goodenough1971} proposed that this mechanism leads to a filled bonding $a_{1g}$ band and a monoclinic insulator \footnote[1]{The notations $d_{||}$ and $\pi^*$ were used for $a_{1g}$ and $e^{\pi}_g$ orbitals respectively in his paper and many other works. }. Although most first principles studies confirm this picture, the resulting band structure cannot explain the 0.6~eV insulating gap \cite{haverkort2005,koethe2006,eguchi2008}. The insulating state was successfully reproduced by combining density functional theory (DFT) \cite{Hohenberg1964,Kohn1965} with cluster dynamical mean field theory (cDMFT) \cite{biermann2005}, suggesting a nontrivial interplay between the lattice distortions and electronic correlations in the M1 phase. Previous DFT+DMFT studies, however, used a wide range of interaction parameters and reached different conclusions regarding the insulating nature of the M1 phase \cite{biermann2005,tomczak2008,lazarovits2010,belozerov2012,weber2012,brito2016}. Also $GW$ calculations \cite{continenza1999,gatti2007} and hybrid functional theory \cite{eyert2011,zhu2012} can reproduce an insulating M1 phase.

Doping, external electric fields, or strain engineering~\cite{sood2021,wegkamp2015,shao2018,liu2018} have been used to explore the MIT in VO$_2$. The observation of a metallic M1 phase after an optical excitation \cite{roach1971} has motivated numerous studies on the underlying mechanism \cite{cavalleri2001,cavalleri2004,kim2006,liu2012,morrison2014,wegkamp2014,laverock2014,jager2017,otto2019,vidas2020,xu2022v}. Two scenarios for the photo-induced transition are conceivable: a structural change from the M1 to the R phase~\cite{cavalleri2004}, or the existence a monoclinic metal (mM) state~\cite{kim2006}. Disentangling the purely electronic from the lattice driven mechanism requires ultra-fast ($<100$~fs) time-resolved techniques~\cite{wegkamp2014}. Evidence of a transient mM phase was found by combining ultrafast electron diffraction with transmissivity measurements~\cite{morrison2014} and time-resolved terahertz spectroscopy~\cite{otto2019}. A quasi-instantaneous gap collapse ($< 50$ fs) was also detected with time-resolved photoelectron spectroscopy~\cite{wegkamp2014} and extreme UV transient absorption spectroscopy~\cite{jager2017}, excluding a transition controlled by the structural dynamics.  

In this Letter, we study the photo-induced dynamics in a realistic model of VO$_2$ using nonequilibrium cDMFT~\cite{aoki2014}. Our calculations with {\it ab initio} derived interaction parameters for the equilibrium M1 phase yield a gap size in agreement with experiments \cite{haverkort2005,koethe2006,eguchi2008}. Using the same setup in nonequilibrium simulations, we demonstrate the existence of a photo-induced metallic state and study its dependence on the laser frequency and polarization. We show that the photo-induced charge transfer from the $a_{1g}$ orbital to the initially empty $e^\pi_g$ orbitals, via an orbital-mixed doublon state, plays an important role in the formation of the mM phase.

\begin{figure}[t]
	\centering
	\includegraphics[width=0.99\linewidth]{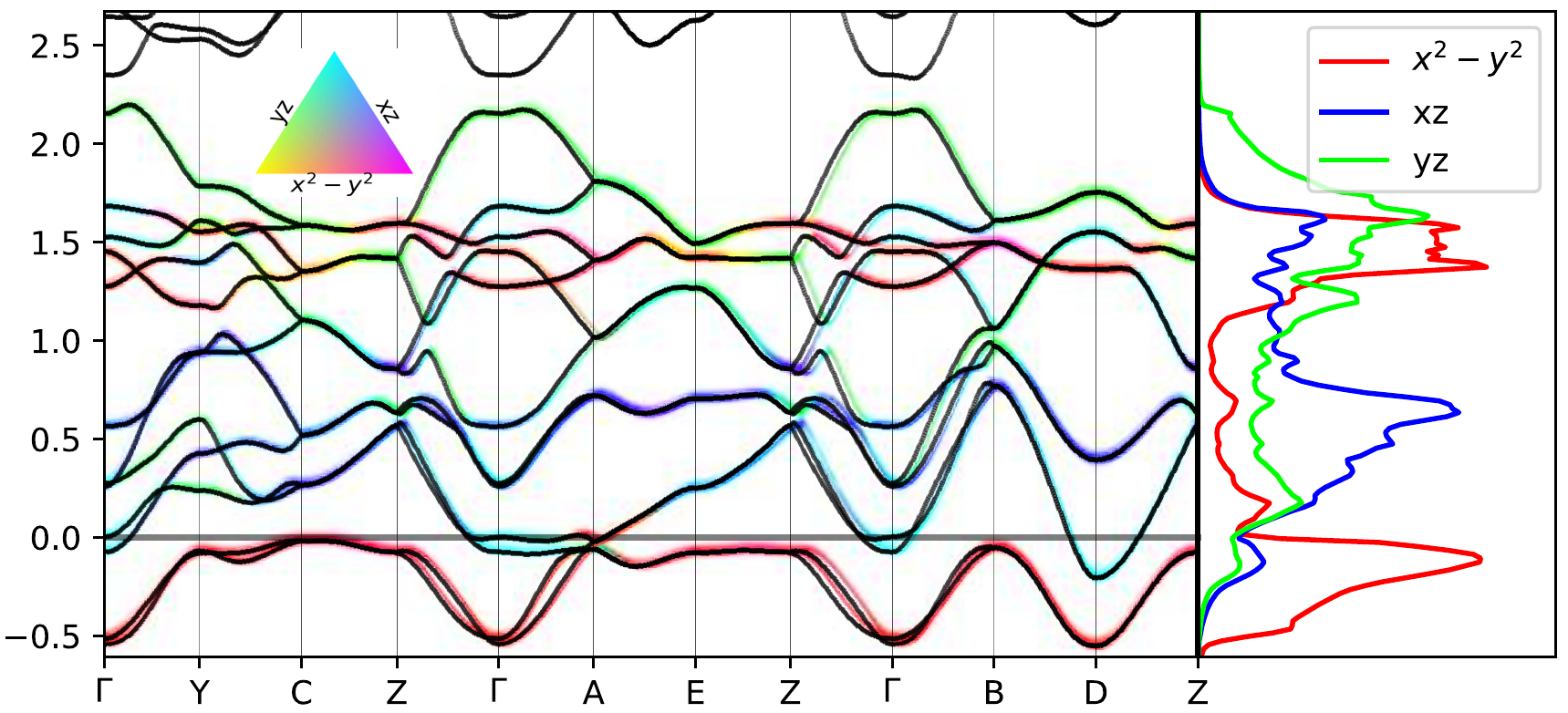}
	\caption{DFT bandstructure (black curves) for VO$_2$ in the M1 phase, and the Wannier interpolated bands with the orbital character indicated by the color. The orbital-projected densities of states are plotted on the right side. 
	}
	\label{fig:bands}
\end{figure}

{\it Model and method.} To derive a realistic model for VO$_2$ in the M1 structure, we start from the experimental lattice structure~\cite{andersson1956,longo1970}, perform DFT calculations using \textsc{Quantum} ESPRESSO \cite{giannozzi2017}, and downfold to the $t_{2g}$ orbitals using Wannier90 \cite{pizzi2020}. The low-energy Hamiltonian at time $t$ is
\begin{equation}
\begin{aligned}\label{eq:hamiltonian}
	\hat{H}(t) =& \sum_{\mathbf{R}}\sum_{ai}\Big\{\sum_{bj} \sum_{\alpha\beta,\sigma} \left[h^{aibj}_{\alpha\beta}(\mathbf{R},t)d_{\alpha \sigma}^{ai\dagger} d^{bj}_{\beta \sigma}+h.c.\right] \nonumber\\
	&\hspace{14mm}-\sum_{\alpha \sigma}\mu n^{ai}_{\alpha \sigma}+H_{\text{K}}^{ai}\Big\},
\end{aligned}
\end{equation}
where $a, b\in \{1,2\}$ label the two V-V dimers in a given unit cell. Within each dimer, $i,j \in \{1,2\}$ are the indices for the two V atoms, $\alpha,\beta \in \{1,2,3\}$ label the three $t_{2g}$ orbitals and $\sigma=\{\uparrow,\downarrow\}$ denotes spin. $n$ is the occupation, $\mu$ the chemical potential and $\mathbf{R}$ labels the unit cell. The interaction term is of the Kanamori type, $H_{\text{K}}^{ai} = \sum_{\alpha} U_\alpha n^{ai}_{\alpha \uparrow} n^{ai}_{\alpha \downarrow} + \sum_{\alpha \neq \beta} U'_{\alpha\beta} n^{ai}_{\alpha \uparrow} n^{ai}_{\beta \downarrow}+ \sum_{\alpha<\beta, \sigma} (U'_{\alpha\beta}-J)n^{ai}_{\alpha \sigma} n^{ai}_{\beta \sigma}-J \sum_{ \alpha \neq \beta} d_{\alpha \uparrow}^{ai\dagger} d^{ai}_{\alpha \downarrow}d_{\beta \downarrow}^{ai\dagger} d^{ai}_{\beta \uparrow}+J \sum_{\alpha \neq \beta} d_{\alpha \uparrow}^{ai\dagger} d^{ai\dagger}_{\alpha \downarrow}d^{ai}_{\beta \downarrow} d^{ai}_{\beta \uparrow}$. Here, $U_\alpha$ is the on-site Coulomb repulsion for orbital $\alpha$, $U^\prime_{\alpha\beta}$ the on-site interaction between different orbitals $\alpha$ and $\beta$, and $J$ the Hund coupling. The interaction parameters are computed with the constrained random-phase approximation (cRPA) \cite{aryasetiawan2004} as implemented in RESPACK \cite{nakamura2021}, yielding the static values $U_{\alpha}=2.2,~2.1$~and~$2.0$~eV for $\alpha=d_{x^2-y^2}$, $d_{xz}$, $d_{yz}$, respectively, and $J=0.28$~eV. The hopping amplitudes $h^{aibj}_{\alpha\beta}(\mathbf{R},t=0)$ extracted from the first principles calculation yield the bandstructure and densities of states (DOS) shown in Fig.~\ref{fig:bands}, which reproduce the DFT results. The effect of the laser pulse is modeled with the Peierls substitution \cite{aoki2014}
\begin{equation}
	h^{aibj}_{\alpha\beta}(\mathbf{R},t) =  h^{aibj}_{\alpha\beta}(\mathbf{R},t=0) e^{-\frac{ie}{\hbar} \phi_{aibj}(\mathbf{R},t)},
\end{equation}
where the Peierls phase for the uniform electric field $\vec{E}(t)$ is $\phi_{aibj}(\mathbf{R},t)= - \int_0^t dt^\prime \vec{E}(t')\cdot(\vec{r}_{bj}-\vec{r}_{ai}+\mathbf{R})$, with $\vec{r}_{ai}$ the position of site $i$ in dimer $a$. The external electric field $\vec{E}(t) = \vec{E}_0\cdot\exp\big(-\frac{(t-t_0)^2}{2\tau^2}\big)\cdot\sin(\omega_0(t-t_0))$ with $\tau = 2.6$~fs (FWHM $6.2$~fs) has a Gaussian envelope, peak amplitude $E_0=0.5$~eV, polarization direction $\hat{E}_0$ and frequency $\omega_0$.
		
To solve the lattice problem, we employ nonequilibrium cDMFT~\cite{eckstein2013,aoki2014,petocchi2023} with a simplified self-consistency \cite{petocchi2023} (see Supplemental Material (SM)) and a noncrossing approximation (NCA) impurity solver~\cite{keiter1971,eckstein2010}. We also analyze an individual dimer using exact diagonalization (ED). The initial temperature is $T=\frac{1}{15}~\text{eV}$. No qualitative changes are expected at lower $T$.  

\begin{figure}[ht]
	\includegraphics[width=0.99\linewidth]{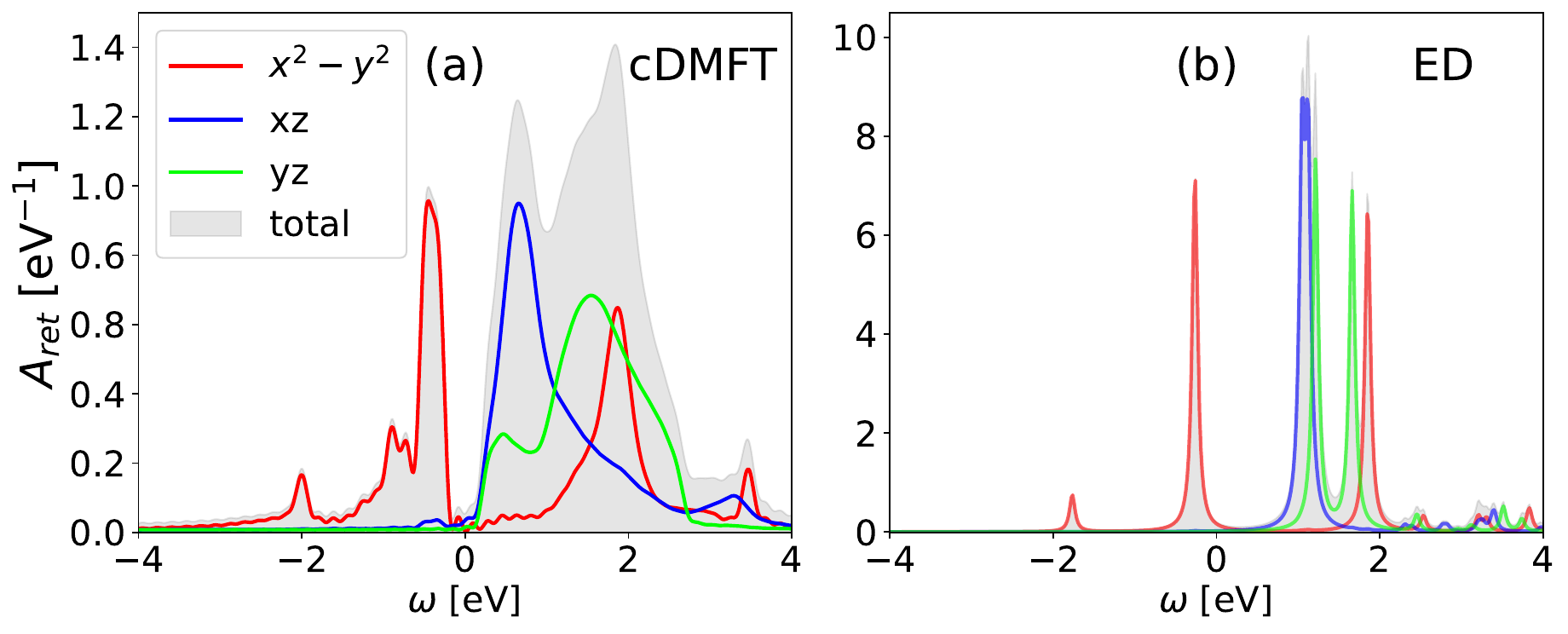}
    \includegraphics[width=0.99\linewidth]{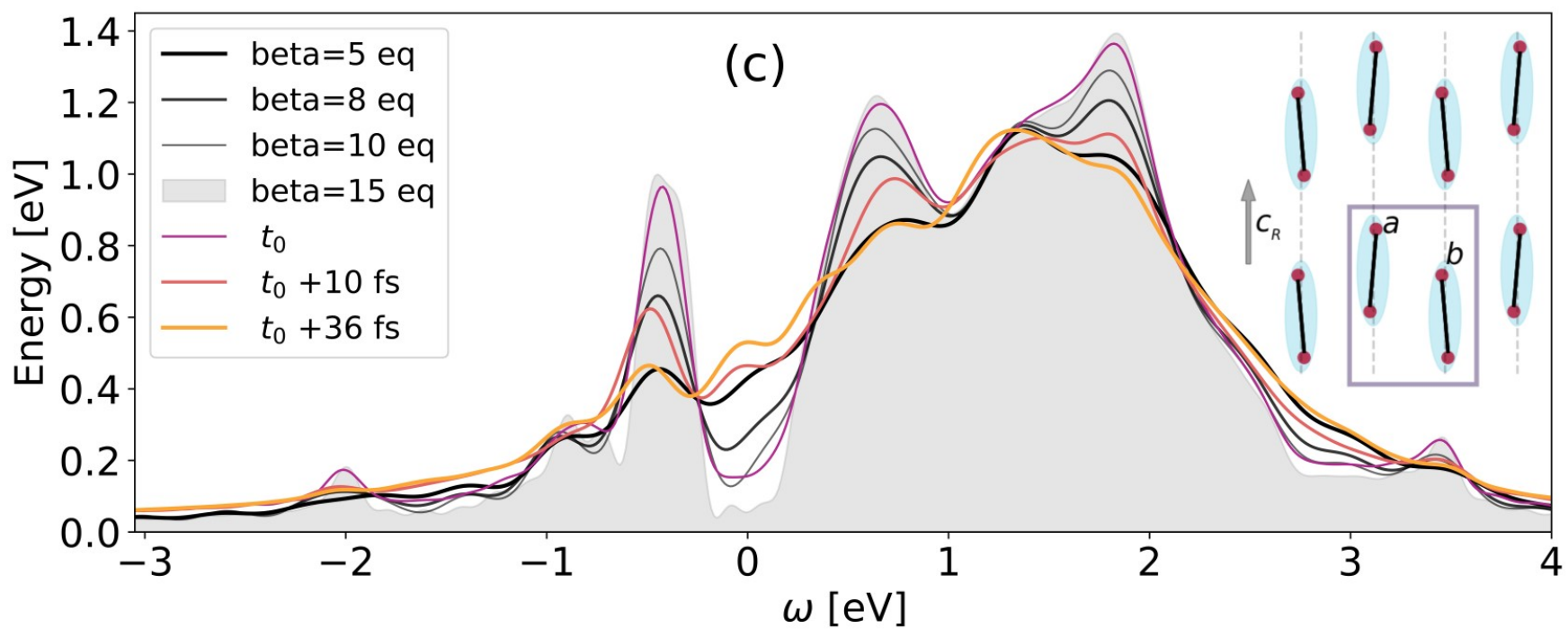}
	\caption{Orbital-resolved and total spectra in equilibrium obtained with cDMFT (a) and ED (b). Both calculations yield a gap between the $d_{x^2-y^2}$ and $d_{xz}$ states. (c) Total spectra for different temperatures (black/gray) and nonequilibrium spectra after the pulse excitation (colored lines). The inset in (c) provides a sketch of the lattice structure. Here, ${\bf c}_R$ indicates the dimerization direction. The unit cell is composed of two dimers with indices $a$ and $b$ (purple box).
	}
	\label{fig:spectra_eq}
\end{figure}

{\it Equilibrium spectrum.}
We first discuss the equilibrium results from cDMFT and ED. As shown in Fig.~\ref{fig:spectra_eq}, the cDMFT spectrum has a gap of 0.5 eV and almost all spectral weight below the Fermi energy is contributed by the $d_{x^2-y^2}$ orbital. The fillings of the $d_{x^2-y^2}$, $d_{xz}$, and $d_{yz}$ orbitals are 0.96, 0.04, and $\sim$\,0 electrons, respectively, in good agreement with previous theoretical \cite{biermann2005,tomczak2008,lazarovits2010,belozerov2012,brito2016} and experimental \cite{haverkort2005,koethe2006,eguchi2008} results. The DOS for the $d_{x^2-y^2}$ orbital features two main peaks at $-0. 45$~eV and $1.86$~eV, and two satellites at $-2$~eV and $3.5$~eV. In the $d_{xz}$ spectrum, only a small feature is located below the Fermi energy, with peak position at $-0.35$~eV, while a prominent peak with a broad high-energy tail exists at $0.65$~eV. The almost empty $d_{yz}$ spectrum exhibits two peaks at $0.46$~eV and $1.54$~eV. 
      
The spectrum of the half-filled Hubbard dimer features two energy levels split by $U$ in the atomic limit or by the bonding-antibonding gap $2h$ ($h$ is the hopping) in the non-interacting limit \cite{petocchi2022}. Based on the ED analysis, we identify the ground state of the realistic V-V dimer as a singlet state with two $d_{x^2-y^2}$ electrons, $|\psi_{\text{GS}}\rangle = 0.89|s\rangle+0.45|d_+ \rangle$, where $|s\rangle=\frac{1}{\sqrt{2}}(|\!\uparrow,\,\downarrow\rangle-|\!\downarrow,\,\uparrow\rangle)$ and $|d_\pm\rangle =\frac{1}{\sqrt{2}} (|\varnothing,\uparrow\downarrow\rangle \pm |\!\downarrow\uparrow,\varnothing\rangle)$, as in the single-orbital Hubbard dimer (SM). Projecting the dimer state of the cDMFT solution onto the singlet state of the $d_{x^2-y^2}$ orbital yields a fidelity of 0.87 in equilibrium.
     
The peaks of the $d_{x^2-y^2}$ spectral function below (above) the Fermi energy correspond to the removal (addition) of an electron from (to) the dimer. In each case, a satellite is split off from the main peak by $\sim 2h_{d_{x^2-y^2}}\approx1.5$~eV (bonding-antibonding splitting, see SM). The main peaks of the $d_{xz}$ orbital in the ED model are  located at $-0.26$~eV (small spectral weight due to the low filling of the orbital) and at $1.05$~eV and $1.12$~eV. The gap size corresponds to the inter-orbital same-spin interaction $U-3J\approx 1.3$~eV, as one would expect for an atomic 3-orbital system with filling $n=1$.  
The peaks of the empty $d_{yz}$ orbital are located at $1.21$~eV and $1.66$~eV, consistent with the bonding-antibonding splitting $2h_{d_{yz}}\approx0.54$~eV. So, in contrast to Ref.~\onlinecite{biermann2005}, the lowest peak above the Fermi energy is associated with the $e^\pi_g$ orbitals, instead of the antibonding $a_{1g}$ orbital, and represents the addition of an electron to the $d_{xz}$ orbital.

This analysis and the comparison between the cDMFT and ED spectra allows us to conclude that the ground state of VO$_2$ in the M1 phase is dominated by singlet states of the $d_{x^2-y^2}$ orbital. The gap in the M1 phase represents a multi-orbital Mott insulating state assisted by the dimerization, in good agreement with experimental results~\cite{koethe2006} and the analysis in Refs.~\onlinecite{lazarovits2010,brito2016}. 

\begin{figure}[t]
	\centering
	\includegraphics[width=0.99\linewidth]{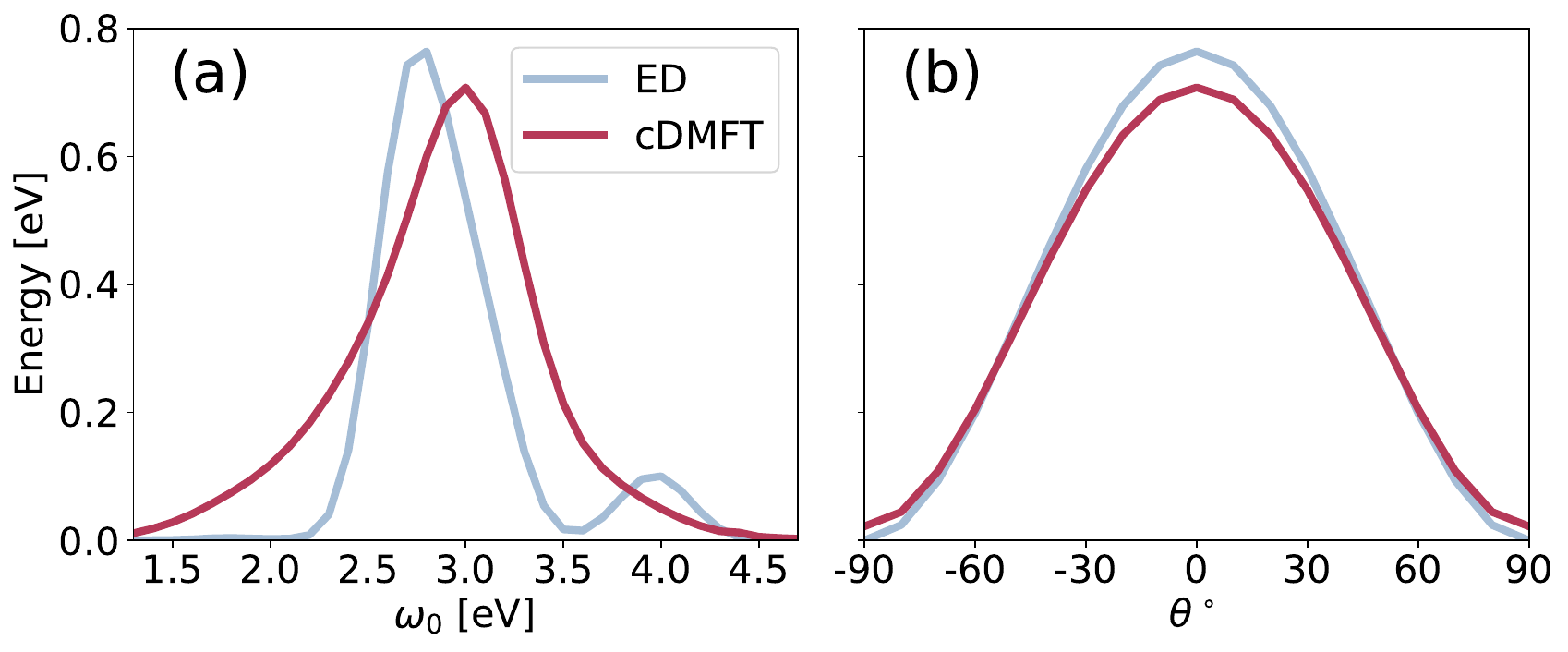}
	\caption{(a) Energy absorption as a function of the pulse frequency $\omega_0$ (left) with $\theta=0^\circ$. (b) Energy absorption as a function of the polarization angle with frequency $\omega_0=3$~eV for cDMFT and $\omega_0=2.8$~eV for ED. $\theta=0^\circ$ corresponds to a polarization along the dimer, in the $\mathbf{c}_R$ direction.
	}
	\label{fig:absorption-frequency}
\end{figure}

{\it Photo-excited system.} 
We next search for the pump frequency $\omega_0$ and polarization $\hat{E}_0=\vec{E_0}/|E_0|$ which yields the maximum energy absorption and study the features of the excited states using both cDMFT and ED calculations.
For this purpose, we tune the polarization angle $\theta$  (the angle between $\hat{E}_0$ and the dimerization axis ${\bf c}_R$, see inset of Fig.~\ref{fig:spectra_eq}(c)) and the frequency $\omega_0$ of the laser pulse. As shown in Fig.~\ref{fig:absorption-frequency}(b), both in the cDMFT and ED simulations, $\theta=0^\circ$ maximizes the absorption. With this polarization fixed, the cDMFT simulations predict the strongest energy absorption for pulse frequency $\omega_0=3.0$~eV, see Fig.~\ref{fig:absorption-frequency}(a).  In the ED analysis, the main absorption peak at $\omega_0=2.8$~eV corresponds to excitations from the ground state $|\psi_{\text{GS}}\rangle$ (singlet state of the $d_{x^2-z^2}$ electrons) to a doublon state $|\psi\rangle = 0.65|d_-\rangle_{xz}+0.76|d_-\rangle_{x^2-y^2}$ with mixed orbital character (henceforth referred to as {\it orbital-mixed doublon state}). This orbital mixture of the photo-doped state is a consequence of the pair hopping term in $H_K$, without which the peak would correspond to the doublon state excitation of a single-orbital Hubbard dimer.
\begin{figure*}[t]
	\centering
    \includegraphics[width=0.88\textwidth]{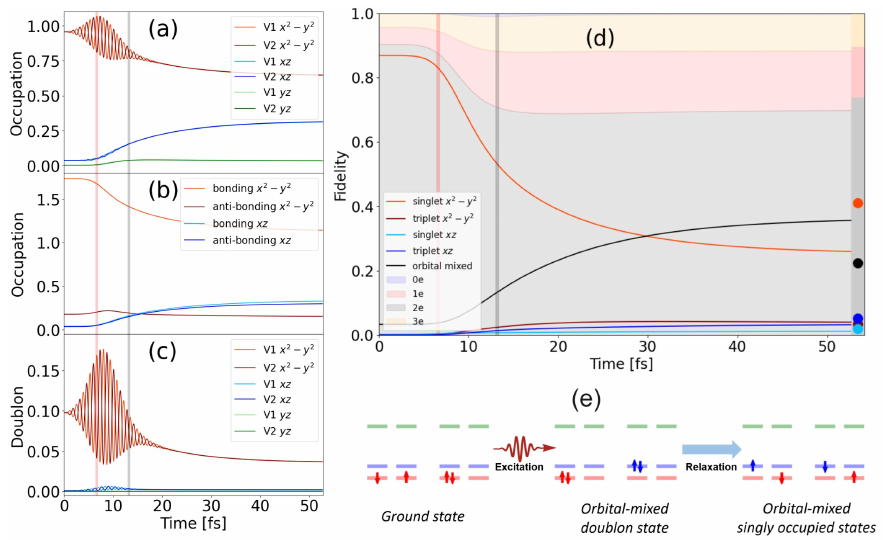}
	\caption{ Occupation in the original site basis (a) and bonding/antibonding basis (b) and doublon density (c) in each orbital of a V-V dimer. (d) Projection of the state occupation of the $d_{x^2-y^2,xz}$ subspace onto electron number sectors (colored shadings). Within the $n=2$ sector of the $d_{x^2-y^2,xz}$ subspace, the states are further projected onto the singlet-triplet basis. The red and gray vertical lines indicate the maximum and the end of the laser pulse, respectively. Panel (e) shows a sketch of the excitation and relaxation pathway. The relevant states are superpositions of the sketched (plus site-permuted) configurations.  			 
	}
	\label{fig:occupation}
\end{figure*}

 In the cDMFT calculations, oscillations in the site occupations and double occupations indicate that the pulse, with maximum at $t=t_0=6.6$~fs (red vertical line), drives the $d_{x^2-y^2}$ electrons between the two dimer sites, see Fig.~\ref{fig:occupation}(a)(c). Since the dimer here is embedded into a lattice environment (mimicked by the cDMFT bath), the injected energy can be converted into various excitations. We observe a rearrangement of charge between the different orbitals and associated with this a reduction of the double occupation in the $d_{x^2-y^2}$ orbital. In particular, as shown in Figure~\ref{fig:occupation}(a), there is a significant flow of charge from the  $d_{x^2-y^2}$ to the  $d_{xz}$ orbitals, which starts before the end of the pulse at $\sim$13~fs (grey vertical line) and persists up to the longest simulation time. The average doublon density decreases, even during the pulse, because of this flow of charge out of the $d_{x^2-y^2}$ orbitals. 
 
 In Fig.~\ref{fig:occupation}(b), we plot for each orbital $\alpha$ the occupation of the bonding 
 \{$\frac{1}{\sqrt{2}}(|\varnothing,\uparrow\rangle-|\!\uparrow,\varnothing\rangle)_\alpha$, $\frac{1}{\sqrt{2}}(|\varnothing,\downarrow\rangle-|\!\downarrow,\varnothing\rangle)_\alpha$\} and antibonding  \{$\frac{1}{\sqrt{2}}(|\varnothing,\!\uparrow\rangle+|\!\uparrow,\varnothing\rangle)_\alpha$, $\frac{1}{\sqrt{2}}(|\varnothing,\downarrow\rangle+|\!\downarrow,\varnothing\rangle)_\alpha$\} states. Most electrons (87\%) are initially in the bonding $d_{x^2-y^2}$ orbital. 
 The laser pulse excites the electrons mostly to the antibonding $d_{x^2-y^2}$ and the $d_{xz}$ orbitals. After the end of the pulse, the electrons in both the bonding and antibonding $d_{x^2-y^2}$ states flow to the $d_{xz}$ orbital, via (strong) pair-hopping and (weak) inter-orbital hopping.
    
 For the following analysis, we define the spin
 singlet states \{$\frac{1}{\sqrt{2}}(|\!\!\uparrow,\downarrow\rangle-|\!\!\downarrow,\uparrow\rangle)_\alpha$, $|\varnothing,\uparrow\downarrow\rangle_\alpha$, $|\!\!\uparrow\downarrow,\varnothing\rangle_\alpha$\} and
 triplet states
 \{$|\!\uparrow,\uparrow\rangle_\alpha$, 
 $\frac{1}{\sqrt{2}}(|\!\uparrow,\downarrow\rangle+|\!\downarrow,\uparrow\rangle)_\alpha$,
 $|\!\downarrow,\downarrow\rangle_\alpha$\} for each of the three orbitals. As shown in Fig.~\ref{fig:occupation}(d), in the equilibrium M1 phase, the sector with $n=2$ electrons, which contains the ground state, has fidelity $>0.9$. The pulse populates mainly states within the $n=2$ sector, but also creates states with $n=1$ and $3$ through charge excitations between the dimers. This charge reshuffling is indicated by the colored shading. After the end of the pulse ($t\gtrsim 13$~fs), the fidelity of the $n=2$ sector slowly increases, and in the absence of energy dissipation (e. g. to phonons) will finally converge to the value corresponding to the thermalized electronic system ($T\sim 2240$~K, calculated from the total energy), which is represented by the color bar on the right. This thermalization takes several hundred fs. 

Within the two-electron sector, the fidelity of the $d_{x^2-y^2}$ singlet decreases rapidly during and after the pulse, while the triplet population in the $d_{x^2-y^2}$ and $d_{xz}$ orbitals increases only slightly. In fact, during and after the excitation, an {\it orbital-mixed state}, with one electron in the $d_{x^2-y^2}$ and the other in the $d_{xz}$ orbital, emerges as the most probable state (black line and Fig.~\ref{fig:occupation}(e)). We present the analogous results obtained with ED in the SM. In the ED analysis, the single dimer is isolated and thus the doublons cannot hop to or exchange energy with other sites, which leads to long-lived oscillations.
 
An important finding is that the population of the orbital-mixed states generates spectral weight in the gap region and is responsible for the almost instantaneous partial gap filling seen in Fig.~\ref{fig:spectra_eq}(c). Our calculations demonstrate how the metallic phase emerging in the photo-doped regime is a consequence of this charge reshuffling between $d_{x^2-y^2}$ and $d_{xz}$ orbitals. We also note that the nonequilibrium spectrum after the pulse is similar to a thermal spectrum corresponding to a high electronic temperature in the M1 structure ($\beta \approx 5 \text{ eV}^{-1}\leftrightarrow T\approx 2200$~K), although it has a higher in-gap population. 

\begin{figure}[t]
	\centering
	\includegraphics[width=0.99\linewidth]{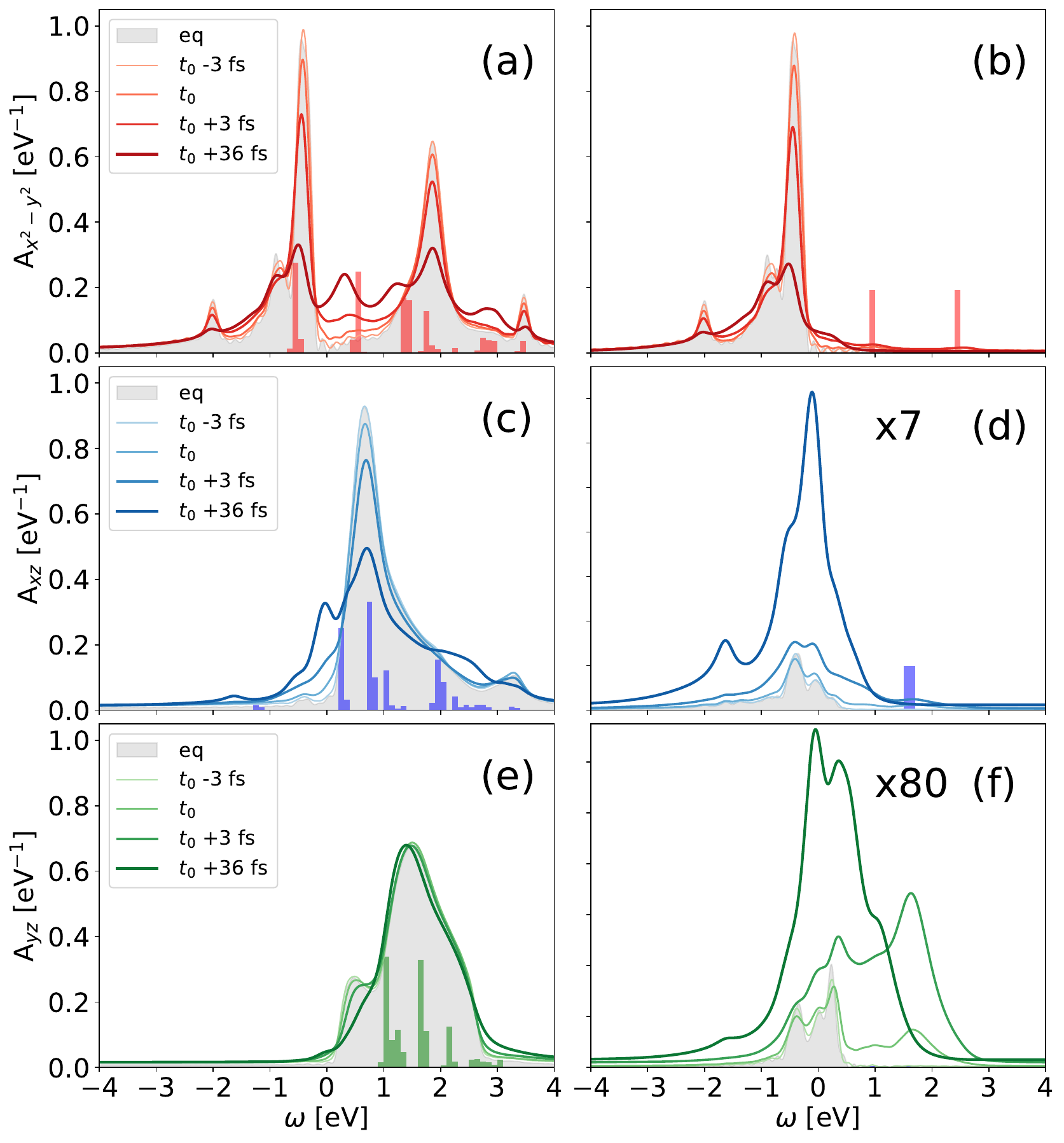}
	\caption{Retarded (left) and lesser (right) spectra for the $d_{x^2-y^2}$ (top), $d_{xz}$ (middle) and $d_{yz}$ (bottom) orbitals obtained with cDMFT. The $\omega_0=3.0$~eV pulse is used. The vertical lines indicate the positions of the poles found in the ED spectra of the orbital-mixed singly occupied states (a,c,e) and the orbital-mixed doublon state (b,d). 
	}
	\label{fig:ppsc_spectra}
\end{figure}
	
We finally present a detailed analysis of the time-dependent cDMFT spectra, obtained for the optimal absorption frequency ($\omega_0=3.0$~eV) and polarization ($\theta=0^\circ$). The orbital-resolved results are shown in Fig.~\ref{fig:ppsc_spectra}, with the left (right) row plotting the total spectral functions $A^{\text{ret}}$ (occupations $A^<$) obtained from the retarded and lesser components of the Keldysh Green's function by the Wigner transformation~\cite{aoki2014}. The photo-excited population exhibits additional peaks above the Fermi energy. In the $d_{x^2-y^2}$ occupations, we observe two short-lived peaks at 2.45~eV and 0.95~eV and in the $d_{xz}$ occupation a peak around 1.6~eV. These transient peaks are visible only around the pump maximum $t\approx t_0$, indicating a rapid decay of the photo-generated states. The vertical lines mark the poles from the ED analysis of the excited state (SM, Fig.~9(e)), which  are in good agreement with  the cDMFT data. The ED results indicate that the transient peaks in $A^<$ correspond to the removal of an electron from the orbital-mixed doublon state, which results in a bonding or anti-bonding state of the $d_{x^2-y^2}$ (two red lines in panel (b)) or $d_{xz}$ (two blue lines in (d)) orbital. The orbital-mixed doublon state is short lived, and decays into {\it orbital-mixed singly occupied states} (Fig.~\ref{fig:occupation}(e)). Hence, during  the photo-doping pulse, orbital-mixed doublons are created, which rapidly decay by transferring an electron to the neighboring unoccupied $d_{x^2-y^2}$ or $d_{xz}$ orbital within the dimer. This dynamics has been observed in experiments as an instantaneous charge transfer effect \cite{morrison2014,wegkamp2014}. Since the resulting orbital-mixed singly occupied states yield spectral weight in the gap (SM Fig.~9 right column) we obtain a transient metallic state in the M1 structure.

{\it Conclusion.} Our  {\it ab initio} nonequilibrium cDMFT simulations of VO$_2$ clarify the photoinduced charge dynamics after photo-excitation with a $\omega_0=3$~eV laser in the M1 phase. The optical excitation induces a quasi-instantaneous gap filling by transiently populating a specific {\it orbital-mixed doublon state}, which rapidly decays into {\it orbital-mixed singly occupied states}, followed by a slow (hundreds of fs) thermalization to a high-$T$ metallic state. On this longer timescale, the lattice is expected to respond and a reliable simulation would require the introduction of phonon degrees of freedom. For the short-time dynamics, our study unambiguously shows that multi-orbital Mott and Hund physics play a key role in the formation of the insulating M1 phase in equilibrium, and in driving electrons into the mM phase after an optical excitation, extending the previously proposed hole-driving mechanism~\cite{wegkamp2014} from the band picture~\cite{he2016,xu2022} to the strongly correlated context appropriate for VO$_2$. The orbital-mixed states generate spectral weight in the gap, while the Mott related features (Hubbard bands) persist. The transient mM phase thus shows a gap filling, but not a gap collapse, similar to what is observed at high electronic temperature. 

{\it Acknowledgments.} This work was supported by the Swiss National Science Foundation via the German Research Unit QUAST (J.C.) and NCCR Marvel (V.C.). The calculations were run on the beo05 cluster at the University of Fribourg, using a code based on NESSi~\cite{schuler2020}.

\bibliography{ref}

\begin{thebibliography}{55}%
\makeatletter
\providecommand \@ifxundefined [1]{%
 \@ifx{#1\undefined}
}%
\providecommand \@ifnum [1]{%
 \ifnum #1\expandafter \@firstoftwo
 \else \expandafter \@secondoftwo
 \fi
}%
\providecommand \@ifx [1]{%
 \ifx #1\expandafter \@firstoftwo
 \else \expandafter \@secondoftwo
 \fi
}%
\providecommand \natexlab [1]{#1}%
\providecommand \enquote  [1]{``#1''}%
\providecommand \bibnamefont  [1]{#1}%
\providecommand \bibfnamefont [1]{#1}%
\providecommand \citenamefont [1]{#1}%
\providecommand \href@noop [0]{\@secondoftwo}%
\providecommand \href [0]{\begingroup \@sanitize@url \@href}%
\providecommand \@href[1]{\@@startlink{#1}\@@href}%
\providecommand \@@href[1]{\endgroup#1\@@endlink}%
\providecommand \@sanitize@url [0]{\catcode `\\12\catcode `\$12\catcode
  `\&12\catcode `\#12\catcode `\^12\catcode `\_12\catcode `\%12\relax}%
\providecommand \@@startlink[1]{}%
\providecommand \@@endlink[0]{}%
\providecommand \url  [0]{\begingroup\@sanitize@url \@url }%
\providecommand \@url [1]{\endgroup\@href {#1}{\urlprefix }}%
\providecommand \urlprefix  [0]{URL }%
\providecommand \Eprint [0]{\href }%
\providecommand \doibase [0]{http://dx.doi.org/}%
\providecommand \selectlanguage [0]{\@gobble}%
\providecommand \bibinfo  [0]{\@secondoftwo}%
\providecommand \bibfield  [0]{\@secondoftwo}%
\providecommand \translation [1]{[#1]}%
\providecommand \BibitemOpen [0]{}%
\providecommand \bibitemStop [0]{}%
\providecommand \bibitemNoStop [0]{.\EOS\space}%
\providecommand \EOS [0]{\spacefactor3000\relax}%
\providecommand \BibitemShut  [1]{\csname bibitem#1\endcsname}%
\let\auto@bib@innerbib\@empty
%</preamble>
\bibitem [{\citenamefont {Morin}(1959)}]{morin1959}%
  \BibitemOpen
  \bibfield  {author} {\bibinfo {author} {\bibfnamefont {F.~J.}\ \bibnamefont
  {Morin}},\ }\href {\doibase 10.1103/PhysRevLett.3.34} {\bibfield  {journal}
  {\bibinfo  {journal} {Phys. Rev. Lett.}\ }\textbf {\bibinfo {volume} {3}},\
  \bibinfo {pages} {34} (\bibinfo {year} {1959})}\BibitemShut {NoStop}%
\bibitem [{\citenamefont {Eyert}(2002)}]{eyert2002}%
  \BibitemOpen
  \bibfield  {author} {\bibinfo {author} {\bibfnamefont {V.}~\bibnamefont
  {Eyert}},\ }\href {\doibase 10.1002/andp.20025140902} {\bibfield  {journal}
  {\bibinfo  {journal} {Annalen der Physik}\ }\textbf {\bibinfo {volume}
  {514}},\ \bibinfo {pages} {650} (\bibinfo {year} {2002})}\BibitemShut
  {NoStop}%
\bibitem [{\citenamefont {Goodenough}(1971)}]{goodenough1971}%
  \BibitemOpen
  \bibfield  {author} {\bibinfo {author} {\bibfnamefont {J.~B.}\ \bibnamefont
  {Goodenough}},\ }\href {\doibase 10.1016/0022-4596(71)90091-0} {\bibfield
  {journal} {\bibinfo  {journal} {Journal of Solid State Chemistry}\ }\textbf
  {\bibinfo {volume} {3}},\ \bibinfo {pages} {490} (\bibinfo {year}
  {1971})}\BibitemShut {NoStop}%
\bibitem [{Note1()}]{Note1}%
  \BibitemOpen
  \bibinfo {note} {The notations $d_{||}$ and $\pi ^*$ were used for $a_{1g}$
  and $e^{\pi }_g$ orbitals respectively in his paper and many other
  works.}\BibitemShut {Stop}%
\bibitem [{\citenamefont {Haverkort}\ \emph {et~al.}(2005)\citenamefont
  {Haverkort}, \citenamefont {Hu}, \citenamefont {Tanaka}, \citenamefont
  {Reichelt}, \citenamefont {Streltsov}, \citenamefont {Korotin}, \citenamefont
  {Anisimov}, \citenamefont {Hsieh}, \citenamefont {Lin}, \citenamefont {Chen},
  \citenamefont {Khomskii},\ and\ \citenamefont {Tjeng}}]{haverkort2005}%
  \BibitemOpen
  \bibfield  {author} {\bibinfo {author} {\bibfnamefont {M.~W.}\ \bibnamefont
  {Haverkort}}, \bibinfo {author} {\bibfnamefont {Z.}~\bibnamefont {Hu}},
  \bibinfo {author} {\bibfnamefont {A.}~\bibnamefont {Tanaka}}, \bibinfo
  {author} {\bibfnamefont {W.}~\bibnamefont {Reichelt}}, \bibinfo {author}
  {\bibfnamefont {S.~V.}\ \bibnamefont {Streltsov}}, \bibinfo {author}
  {\bibfnamefont {M.~A.}\ \bibnamefont {Korotin}}, \bibinfo {author}
  {\bibfnamefont {V.~I.}\ \bibnamefont {Anisimov}}, \bibinfo {author}
  {\bibfnamefont {H.~H.}\ \bibnamefont {Hsieh}}, \bibinfo {author}
  {\bibfnamefont {H.-J.}\ \bibnamefont {Lin}}, \bibinfo {author} {\bibfnamefont
  {C.~T.}\ \bibnamefont {Chen}}, \bibinfo {author} {\bibfnamefont {D.~I.}\
  \bibnamefont {Khomskii}}, \ and\ \bibinfo {author} {\bibfnamefont {L.~H.}\
  \bibnamefont {Tjeng}},\ }\href {\doibase 10.1103/PhysRevLett.95.196404}
  {\bibfield  {journal} {\bibinfo  {journal} {Phys. Rev. Lett.}\ }\textbf
  {\bibinfo {volume} {95}},\ \bibinfo {pages} {196404} (\bibinfo {year}
  {2005})}\BibitemShut {NoStop}%
\bibitem [{\citenamefont {Koethe}\ \emph {et~al.}(2006)\citenamefont {Koethe},
  \citenamefont {Hu}, \citenamefont {Haverkort}, \citenamefont
  {{Sch{\"u}{\ss}ler-Langeheine}}, \citenamefont {Venturini}, \citenamefont
  {Brookes}, \citenamefont {Tjernberg}, \citenamefont {Reichelt}, \citenamefont
  {Hsieh}, \citenamefont {Lin}, \citenamefont {Chen},\ and\ \citenamefont
  {Tjeng}}]{koethe2006}%
  \BibitemOpen
  \bibfield  {author} {\bibinfo {author} {\bibfnamefont {T.~C.}\ \bibnamefont
  {Koethe}}, \bibinfo {author} {\bibfnamefont {Z.}~\bibnamefont {Hu}}, \bibinfo
  {author} {\bibfnamefont {M.~W.}\ \bibnamefont {Haverkort}}, \bibinfo {author}
  {\bibfnamefont {C.}~\bibnamefont {{Sch{\"u}{\ss}ler-Langeheine}}}, \bibinfo
  {author} {\bibfnamefont {F.}~\bibnamefont {Venturini}}, \bibinfo {author}
  {\bibfnamefont {N.~B.}\ \bibnamefont {Brookes}}, \bibinfo {author}
  {\bibfnamefont {O.}~\bibnamefont {Tjernberg}}, \bibinfo {author}
  {\bibfnamefont {W.}~\bibnamefont {Reichelt}}, \bibinfo {author}
  {\bibfnamefont {H.~H.}\ \bibnamefont {Hsieh}}, \bibinfo {author}
  {\bibfnamefont {H.-J.}\ \bibnamefont {Lin}}, \bibinfo {author} {\bibfnamefont
  {C.~T.}\ \bibnamefont {Chen}}, \ and\ \bibinfo {author} {\bibfnamefont
  {L.~H.}\ \bibnamefont {Tjeng}},\ }\href {\doibase
  10.1103/PhysRevLett.97.116402} {\bibfield  {journal} {\bibinfo  {journal}
  {Phys. Rev. Lett.}\ }\textbf {\bibinfo {volume} {97}},\ \bibinfo {pages}
  {116402} (\bibinfo {year} {2006})}\BibitemShut {NoStop}%
\bibitem [{\citenamefont {Eguchi}\ \emph {et~al.}(2008)\citenamefont {Eguchi},
  \citenamefont {Taguchi}, \citenamefont {Matsunami}, \citenamefont {Horiba},
  \citenamefont {Yamamoto}, \citenamefont {Ishida}, \citenamefont {Chainani},
  \citenamefont {Takata}, \citenamefont {Yabashi}, \citenamefont {Miwa},
  \citenamefont {Nishino}, \citenamefont {Tamasaku}, \citenamefont {Ishikawa},
  \citenamefont {Senba}, \citenamefont {Ohashi}, \citenamefont {Muraoka},
  \citenamefont {Hiroi},\ and\ \citenamefont {Shin}}]{eguchi2008}%
  \BibitemOpen
  \bibfield  {author} {\bibinfo {author} {\bibfnamefont {R.}~\bibnamefont
  {Eguchi}}, \bibinfo {author} {\bibfnamefont {M.}~\bibnamefont {Taguchi}},
  \bibinfo {author} {\bibfnamefont {M.}~\bibnamefont {Matsunami}}, \bibinfo
  {author} {\bibfnamefont {K.}~\bibnamefont {Horiba}}, \bibinfo {author}
  {\bibfnamefont {K.}~\bibnamefont {Yamamoto}}, \bibinfo {author}
  {\bibfnamefont {Y.}~\bibnamefont {Ishida}}, \bibinfo {author} {\bibfnamefont
  {A.}~\bibnamefont {Chainani}}, \bibinfo {author} {\bibfnamefont
  {Y.}~\bibnamefont {Takata}}, \bibinfo {author} {\bibfnamefont
  {M.}~\bibnamefont {Yabashi}}, \bibinfo {author} {\bibfnamefont
  {D.}~\bibnamefont {Miwa}}, \bibinfo {author} {\bibfnamefont {Y.}~\bibnamefont
  {Nishino}}, \bibinfo {author} {\bibfnamefont {K.}~\bibnamefont {Tamasaku}},
  \bibinfo {author} {\bibfnamefont {T.}~\bibnamefont {Ishikawa}}, \bibinfo
  {author} {\bibfnamefont {Y.}~\bibnamefont {Senba}}, \bibinfo {author}
  {\bibfnamefont {H.}~\bibnamefont {Ohashi}}, \bibinfo {author} {\bibfnamefont
  {Y.}~\bibnamefont {Muraoka}}, \bibinfo {author} {\bibfnamefont
  {Z.}~\bibnamefont {Hiroi}}, \ and\ \bibinfo {author} {\bibfnamefont
  {S.}~\bibnamefont {Shin}},\ }\href {\doibase 10.1103/PhysRevB.78.075115}
  {\bibfield  {journal} {\bibinfo  {journal} {Phys. Rev. B}\ }\textbf {\bibinfo
  {volume} {78}},\ \bibinfo {pages} {075115} (\bibinfo {year}
  {2008})}\BibitemShut {NoStop}%
\bibitem [{\citenamefont {Hohenberg}\ and\ \citenamefont
  {Kohn}(1964)}]{Hohenberg1964}%
  \BibitemOpen
  \bibfield  {author} {\bibinfo {author} {\bibfnamefont {P.}~\bibnamefont
  {Hohenberg}}\ and\ \bibinfo {author} {\bibfnamefont {W.}~\bibnamefont
  {Kohn}},\ }\href {\doibase 10.1103/PhysRev.136.B864} {\bibfield  {journal}
  {\bibinfo  {journal} {Phys. Rev.}\ }\textbf {\bibinfo {volume} {136}},\
  \bibinfo {pages} {B864} (\bibinfo {year} {1964})}\BibitemShut {NoStop}%
\bibitem [{\citenamefont {Kohn}\ and\ \citenamefont {Sham}(1965)}]{Kohn1965}%
  \BibitemOpen
  \bibfield  {author} {\bibinfo {author} {\bibfnamefont {W.}~\bibnamefont
  {Kohn}}\ and\ \bibinfo {author} {\bibfnamefont {L.~J.}\ \bibnamefont
  {Sham}},\ }\href {\doibase 10.1103/PhysRev.140.A1133} {\bibfield  {journal}
  {\bibinfo  {journal} {Phys. Rev.}\ }\textbf {\bibinfo {volume} {140}},\
  \bibinfo {pages} {A1133} (\bibinfo {year} {1965})}\BibitemShut {NoStop}%
\bibitem [{\citenamefont {Biermann}\ \emph {et~al.}(2005)\citenamefont
  {Biermann}, \citenamefont {Poteryaev}, \citenamefont {Lichtenstein},\ and\
  \citenamefont {Georges}}]{biermann2005}%
  \BibitemOpen
  \bibfield  {author} {\bibinfo {author} {\bibfnamefont {S.}~\bibnamefont
  {Biermann}}, \bibinfo {author} {\bibfnamefont {A.}~\bibnamefont {Poteryaev}},
  \bibinfo {author} {\bibfnamefont {A.~I.}\ \bibnamefont {Lichtenstein}}, \
  and\ \bibinfo {author} {\bibfnamefont {A.}~\bibnamefont {Georges}},\ }\href
  {\doibase 10.1103/PhysRevLett.94.026404} {\bibfield  {journal} {\bibinfo
  {journal} {Phys. Rev. Lett.}\ }\textbf {\bibinfo {volume} {94}},\ \bibinfo
  {pages} {026404} (\bibinfo {year} {2005})}\BibitemShut {NoStop}%
\bibitem [{\citenamefont {Tomczak}\ \emph {et~al.}(2008)\citenamefont
  {Tomczak}, \citenamefont {Aryasetiawan},\ and\ \citenamefont
  {Biermann}}]{tomczak2008}%
  \BibitemOpen
  \bibfield  {author} {\bibinfo {author} {\bibfnamefont {J.~M.}\ \bibnamefont
  {Tomczak}}, \bibinfo {author} {\bibfnamefont {F.}~\bibnamefont
  {Aryasetiawan}}, \ and\ \bibinfo {author} {\bibfnamefont {S.}~\bibnamefont
  {Biermann}},\ }\href {\doibase 10.1103/PhysRevB.78.115103} {\bibfield
  {journal} {\bibinfo  {journal} {Phys. Rev. B}\ }\textbf {\bibinfo {volume}
  {78}},\ \bibinfo {pages} {115103} (\bibinfo {year} {2008})}\BibitemShut
  {NoStop}%
\bibitem [{\citenamefont {Lazarovits}\ \emph {et~al.}(2010)\citenamefont
  {Lazarovits}, \citenamefont {Kim}, \citenamefont {Haule},\ and\ \citenamefont
  {Kotliar}}]{lazarovits2010}%
  \BibitemOpen
  \bibfield  {author} {\bibinfo {author} {\bibfnamefont {B.}~\bibnamefont
  {Lazarovits}}, \bibinfo {author} {\bibfnamefont {K.}~\bibnamefont {Kim}},
  \bibinfo {author} {\bibfnamefont {K.}~\bibnamefont {Haule}}, \ and\ \bibinfo
  {author} {\bibfnamefont {G.}~\bibnamefont {Kotliar}},\ }\href {\doibase
  10.1103/PhysRevB.81.115117} {\bibfield  {journal} {\bibinfo  {journal} {Phys.
  Rev. B}\ }\textbf {\bibinfo {volume} {81}},\ \bibinfo {pages} {115117}
  (\bibinfo {year} {2010})}\BibitemShut {NoStop}%
\bibitem [{\citenamefont {Belozerov}\ \emph {et~al.}(2012)\citenamefont
  {Belozerov}, \citenamefont {Korotin}, \citenamefont {Anisimov},\ and\
  \citenamefont {Poteryaev}}]{belozerov2012}%
  \BibitemOpen
  \bibfield  {author} {\bibinfo {author} {\bibfnamefont {A.~S.}\ \bibnamefont
  {Belozerov}}, \bibinfo {author} {\bibfnamefont {M.~A.}\ \bibnamefont
  {Korotin}}, \bibinfo {author} {\bibfnamefont {V.~I.}\ \bibnamefont
  {Anisimov}}, \ and\ \bibinfo {author} {\bibfnamefont {A.~I.}\ \bibnamefont
  {Poteryaev}},\ }\href {\doibase 10.1103/PhysRevB.85.045109} {\bibfield
  {journal} {\bibinfo  {journal} {Phys. Rev. B}\ }\textbf {\bibinfo {volume}
  {85}},\ \bibinfo {pages} {045109} (\bibinfo {year} {2012})}\BibitemShut
  {NoStop}%
\bibitem [{\citenamefont {Weber}\ \emph {et~al.}(2012)\citenamefont {Weber},
  \citenamefont {O'Regan}, \citenamefont {Hine}, \citenamefont {Payne},
  \citenamefont {Kotliar},\ and\ \citenamefont {Littlewood}}]{weber2012}%
  \BibitemOpen
  \bibfield  {author} {\bibinfo {author} {\bibfnamefont {C.}~\bibnamefont
  {Weber}}, \bibinfo {author} {\bibfnamefont {D.~D.}\ \bibnamefont {O'Regan}},
  \bibinfo {author} {\bibfnamefont {N.~D.~M.}\ \bibnamefont {Hine}}, \bibinfo
  {author} {\bibfnamefont {M.~C.}\ \bibnamefont {Payne}}, \bibinfo {author}
  {\bibfnamefont {G.}~\bibnamefont {Kotliar}}, \ and\ \bibinfo {author}
  {\bibfnamefont {P.~B.}\ \bibnamefont {Littlewood}},\ }\href {\doibase
  10.1103/PhysRevLett.108.256402} {\bibfield  {journal} {\bibinfo  {journal}
  {Phys. Rev. Lett.}\ }\textbf {\bibinfo {volume} {108}},\ \bibinfo {pages}
  {256402} (\bibinfo {year} {2012})}\BibitemShut {NoStop}%
\bibitem [{\citenamefont {Brito}\ \emph {et~al.}(2016)\citenamefont {Brito},
  \citenamefont {Aguiar}, \citenamefont {Haule},\ and\ \citenamefont
  {Kotliar}}]{brito2016}%
  \BibitemOpen
  \bibfield  {author} {\bibinfo {author} {\bibfnamefont {W.~H.}\ \bibnamefont
  {Brito}}, \bibinfo {author} {\bibfnamefont {M.~C.~O.}\ \bibnamefont
  {Aguiar}}, \bibinfo {author} {\bibfnamefont {K.}~\bibnamefont {Haule}}, \
  and\ \bibinfo {author} {\bibfnamefont {G.}~\bibnamefont {Kotliar}},\ }\href
  {\doibase 10.1103/PhysRevLett.117.056402} {\bibfield  {journal} {\bibinfo
  {journal} {Phys. Rev. Lett.}\ }\textbf {\bibinfo {volume} {117}},\ \bibinfo
  {pages} {056402} (\bibinfo {year} {2016})}\BibitemShut {NoStop}%
\bibitem [{\citenamefont {Continenza}\ \emph {et~al.}(1999)\citenamefont
  {Continenza}, \citenamefont {Massidda},\ and\ \citenamefont
  {Posternak}}]{continenza1999}%
  \BibitemOpen
  \bibfield  {author} {\bibinfo {author} {\bibfnamefont {A.}~\bibnamefont
  {Continenza}}, \bibinfo {author} {\bibfnamefont {S.}~\bibnamefont
  {Massidda}}, \ and\ \bibinfo {author} {\bibfnamefont {M.}~\bibnamefont
  {Posternak}},\ }\href {\doibase 10.1103/PhysRevB.60.15699} {\bibfield
  {journal} {\bibinfo  {journal} {Phys. Rev. B}\ }\textbf {\bibinfo {volume}
  {60}},\ \bibinfo {pages} {15699} (\bibinfo {year} {1999})}\BibitemShut
  {NoStop}%
\bibitem [{\citenamefont {Gatti}\ \emph {et~al.}(2007)\citenamefont {Gatti},
  \citenamefont {Bruneval}, \citenamefont {Olevano},\ and\ \citenamefont
  {Reining}}]{gatti2007}%
  \BibitemOpen
  \bibfield  {author} {\bibinfo {author} {\bibfnamefont {M.}~\bibnamefont
  {Gatti}}, \bibinfo {author} {\bibfnamefont {F.}~\bibnamefont {Bruneval}},
  \bibinfo {author} {\bibfnamefont {V.}~\bibnamefont {Olevano}}, \ and\
  \bibinfo {author} {\bibfnamefont {L.}~\bibnamefont {Reining}},\ }\href
  {\doibase 10.1103/PhysRevLett.99.266402} {\bibfield  {journal} {\bibinfo
  {journal} {Phys. Rev. Lett.}\ }\textbf {\bibinfo {volume} {99}},\ \bibinfo
  {pages} {266402} (\bibinfo {year} {2007})}\BibitemShut {NoStop}%
\bibitem [{\citenamefont {Eyert}(2011)}]{eyert2011}%
  \BibitemOpen
  \bibfield  {author} {\bibinfo {author} {\bibfnamefont {V.}~\bibnamefont
  {Eyert}},\ }\href {\doibase 10.1103/PhysRevLett.107.016401} {\bibfield
  {journal} {\bibinfo  {journal} {Phys. Rev. Lett.}\ }\textbf {\bibinfo
  {volume} {107}},\ \bibinfo {pages} {016401} (\bibinfo {year}
  {2011})}\BibitemShut {NoStop}%
\bibitem [{\citenamefont {Zhu}\ and\ \citenamefont
  {Schwingenschl\"ogl}(2012)}]{zhu2012}%
  \BibitemOpen
  \bibfield  {author} {\bibinfo {author} {\bibfnamefont {Z.}~\bibnamefont
  {Zhu}}\ and\ \bibinfo {author} {\bibfnamefont {U.}~\bibnamefont
  {Schwingenschl\"ogl}},\ }\href {\doibase 10.1103/PhysRevB.86.075149}
  {\bibfield  {journal} {\bibinfo  {journal} {Phys. Rev. B}\ }\textbf {\bibinfo
  {volume} {86}},\ \bibinfo {pages} {075149} (\bibinfo {year}
  {2012})}\BibitemShut {NoStop}%
\bibitem [{\citenamefont {Sood}\ \emph {et~al.}(2021)\citenamefont {Sood},
  \citenamefont {Shen}, \citenamefont {Shi}, \citenamefont {Kumar},
  \citenamefont {Park}, \citenamefont {Zajac}, \citenamefont {Sun},
  \citenamefont {Chen}, \citenamefont {Ramanathan}, \citenamefont {Wang},
  \citenamefont {Chueh},\ and\ \citenamefont {Lindenberg}}]{sood2021}%
  \BibitemOpen
  \bibfield  {author} {\bibinfo {author} {\bibfnamefont {A.}~\bibnamefont
  {Sood}}, \bibinfo {author} {\bibfnamefont {X.}~\bibnamefont {Shen}}, \bibinfo
  {author} {\bibfnamefont {Y.}~\bibnamefont {Shi}}, \bibinfo {author}
  {\bibfnamefont {S.}~\bibnamefont {Kumar}}, \bibinfo {author} {\bibfnamefont
  {S.~J.}\ \bibnamefont {Park}}, \bibinfo {author} {\bibfnamefont
  {M.}~\bibnamefont {Zajac}}, \bibinfo {author} {\bibfnamefont
  {Y.}~\bibnamefont {Sun}}, \bibinfo {author} {\bibfnamefont {L.-Q.}\
  \bibnamefont {Chen}}, \bibinfo {author} {\bibfnamefont {S.}~\bibnamefont
  {Ramanathan}}, \bibinfo {author} {\bibfnamefont {X.}~\bibnamefont {Wang}},
  \bibinfo {author} {\bibfnamefont {W.~C.}\ \bibnamefont {Chueh}}, \ and\
  \bibinfo {author} {\bibfnamefont {A.~M.}\ \bibnamefont {Lindenberg}},\ }\href
  {\doibase 10.1126/science.abc0652} {\bibfield  {journal} {\bibinfo  {journal}
  {Science}\ }\textbf {\bibinfo {volume} {373}},\ \bibinfo {pages} {352}
  (\bibinfo {year} {2021})}\BibitemShut {NoStop}%
\bibitem [{\citenamefont {Wegkamp}\ and\ \citenamefont
  {St{\"a}hler}(2015)}]{wegkamp2015}%
  \BibitemOpen
  \bibfield  {author} {\bibinfo {author} {\bibfnamefont {D.}~\bibnamefont
  {Wegkamp}}\ and\ \bibinfo {author} {\bibfnamefont {J.}~\bibnamefont
  {St{\"a}hler}},\ }\href {\doibase 10.1016/j.progsurf.2015.10.001} {\bibfield
  {journal} {\bibinfo  {journal} {Progress in Surface Science}\ }\textbf
  {\bibinfo {volume} {90}},\ \bibinfo {pages} {464} (\bibinfo {year}
  {2015})}\BibitemShut {NoStop}%
\bibitem [{\citenamefont {Shao}\ \emph {et~al.}(2018)\citenamefont {Shao},
  \citenamefont {Cao}, \citenamefont {Luo},\ and\ \citenamefont
  {Jin}}]{shao2018}%
  \BibitemOpen
  \bibfield  {author} {\bibinfo {author} {\bibfnamefont {Z.}~\bibnamefont
  {Shao}}, \bibinfo {author} {\bibfnamefont {X.}~\bibnamefont {Cao}}, \bibinfo
  {author} {\bibfnamefont {H.}~\bibnamefont {Luo}}, \ and\ \bibinfo {author}
  {\bibfnamefont {P.}~\bibnamefont {Jin}},\ }\href {\doibase
  10.1038/s41427-018-0061-2} {\bibfield  {journal} {\bibinfo  {journal} {NPG
  Asia Mater}\ }\textbf {\bibinfo {volume} {10}},\ \bibinfo {pages} {581}
  (\bibinfo {year} {2018})}\BibitemShut {NoStop}%
\bibitem [{\citenamefont {Liu}\ \emph {et~al.}(2018)\citenamefont {Liu},
  \citenamefont {Lee}, \citenamefont {Yang}, \citenamefont {Delaire},\ and\
  \citenamefont {Wu}}]{liu2018}%
  \BibitemOpen
  \bibfield  {author} {\bibinfo {author} {\bibfnamefont {K.}~\bibnamefont
  {Liu}}, \bibinfo {author} {\bibfnamefont {S.}~\bibnamefont {Lee}}, \bibinfo
  {author} {\bibfnamefont {S.}~\bibnamefont {Yang}}, \bibinfo {author}
  {\bibfnamefont {O.}~\bibnamefont {Delaire}}, \ and\ \bibinfo {author}
  {\bibfnamefont {J.}~\bibnamefont {Wu}},\ }\href {\doibase
  10.1016/j.mattod.2018.03.029} {\bibfield  {journal} {\bibinfo  {journal}
  {Materials Today}\ }\textbf {\bibinfo {volume} {21}},\ \bibinfo {pages} {875}
  (\bibinfo {year} {2018})}\BibitemShut {NoStop}%
\bibitem [{\citenamefont {Roach}\ and\ \citenamefont
  {Balberg}(1971)}]{roach1971}%
  \BibitemOpen
  \bibfield  {author} {\bibinfo {author} {\bibfnamefont {W.~R.}\ \bibnamefont
  {Roach}}\ and\ \bibinfo {author} {\bibfnamefont {I.}~\bibnamefont
  {Balberg}},\ }\href {\doibase 10.1016/0038-1098(71)90144-X} {\bibfield
  {journal} {\bibinfo  {journal} {Solid State Communications}\ }\textbf
  {\bibinfo {volume} {9}},\ \bibinfo {pages} {551} (\bibinfo {year}
  {1971})}\BibitemShut {NoStop}%
\bibitem [{\citenamefont {Cavalleri}\ \emph {et~al.}(2001)\citenamefont
  {Cavalleri}, \citenamefont {T\'oth}, \citenamefont {Siders}, \citenamefont
  {Squier}, \citenamefont {R\'aksi}, \citenamefont {Forget},\ and\
  \citenamefont {Kieffer}}]{cavalleri2001}%
  \BibitemOpen
  \bibfield  {author} {\bibinfo {author} {\bibfnamefont {A.}~\bibnamefont
  {Cavalleri}}, \bibinfo {author} {\bibfnamefont {C.}~\bibnamefont {T\'oth}},
  \bibinfo {author} {\bibfnamefont {C.~W.}\ \bibnamefont {Siders}}, \bibinfo
  {author} {\bibfnamefont {J.~A.}\ \bibnamefont {Squier}}, \bibinfo {author}
  {\bibfnamefont {F.}~\bibnamefont {R\'aksi}}, \bibinfo {author} {\bibfnamefont
  {P.}~\bibnamefont {Forget}}, \ and\ \bibinfo {author} {\bibfnamefont {J.~C.}\
  \bibnamefont {Kieffer}},\ }\href {\doibase 10.1103/PhysRevLett.87.237401}
  {\bibfield  {journal} {\bibinfo  {journal} {Phys. Rev. Lett.}\ }\textbf
  {\bibinfo {volume} {87}},\ \bibinfo {pages} {237401} (\bibinfo {year}
  {2001})}\BibitemShut {NoStop}%
\bibitem [{\citenamefont {Cavalleri}\ \emph {et~al.}(2004)\citenamefont
  {Cavalleri}, \citenamefont {Dekorsy}, \citenamefont {Chong}, \citenamefont
  {Kieffer},\ and\ \citenamefont {Schoenlein}}]{cavalleri2004}%
  \BibitemOpen
  \bibfield  {author} {\bibinfo {author} {\bibfnamefont {A.}~\bibnamefont
  {Cavalleri}}, \bibinfo {author} {\bibfnamefont {{\relax Th}.}~\bibnamefont
  {Dekorsy}}, \bibinfo {author} {\bibfnamefont {H.~H.~W.}\ \bibnamefont
  {Chong}}, \bibinfo {author} {\bibfnamefont {J.~C.}\ \bibnamefont {Kieffer}},
  \ and\ \bibinfo {author} {\bibfnamefont {R.~W.}\ \bibnamefont {Schoenlein}},\
  }\href {\doibase 10.1103/PhysRevB.70.161102} {\bibfield  {journal} {\bibinfo
  {journal} {Phys. Rev. B}\ }\textbf {\bibinfo {volume} {70}},\ \bibinfo
  {pages} {161102} (\bibinfo {year} {2004})}\BibitemShut {NoStop}%
\bibitem [{\citenamefont {Kim}\ \emph {et~al.}(2006)\citenamefont {Kim},
  \citenamefont {Lee}, \citenamefont {Kim}, \citenamefont {Chae}, \citenamefont
  {Yun}, \citenamefont {Kang}, \citenamefont {Han}, \citenamefont {Yee},\ and\
  \citenamefont {Lim}}]{kim2006}%
  \BibitemOpen
  \bibfield  {author} {\bibinfo {author} {\bibfnamefont {H.-T.}\ \bibnamefont
  {Kim}}, \bibinfo {author} {\bibfnamefont {Y.~W.}\ \bibnamefont {Lee}},
  \bibinfo {author} {\bibfnamefont {B.-J.}\ \bibnamefont {Kim}}, \bibinfo
  {author} {\bibfnamefont {B.-G.}\ \bibnamefont {Chae}}, \bibinfo {author}
  {\bibfnamefont {S.~J.}\ \bibnamefont {Yun}}, \bibinfo {author} {\bibfnamefont
  {K.-Y.}\ \bibnamefont {Kang}}, \bibinfo {author} {\bibfnamefont {K.-J.}\
  \bibnamefont {Han}}, \bibinfo {author} {\bibfnamefont {K.-J.}\ \bibnamefont
  {Yee}}, \ and\ \bibinfo {author} {\bibfnamefont {Y.-S.}\ \bibnamefont
  {Lim}},\ }\href {\doibase 10.1103/PhysRevLett.97.266401} {\bibfield
  {journal} {\bibinfo  {journal} {Phys. Rev. Lett.}\ }\textbf {\bibinfo
  {volume} {97}},\ \bibinfo {pages} {266401} (\bibinfo {year}
  {2006})}\BibitemShut {NoStop}%
\bibitem [{\citenamefont {Liu}\ \emph {et~al.}(2012)\citenamefont {Liu},
  \citenamefont {Hwang}, \citenamefont {Tao}, \citenamefont {Strikwerda},
  \citenamefont {Fan}, \citenamefont {Keiser}, \citenamefont {Sternbach},
  \citenamefont {West}, \citenamefont {Kittiwatanakul}, \citenamefont {Lu},
  \citenamefont {Wolf}, \citenamefont {Omenetto}, \citenamefont {Zhang},
  \citenamefont {Nelson},\ and\ \citenamefont {Averitt}}]{liu2012}%
  \BibitemOpen
  \bibfield  {author} {\bibinfo {author} {\bibfnamefont {M.}~\bibnamefont
  {Liu}}, \bibinfo {author} {\bibfnamefont {H.~Y.}\ \bibnamefont {Hwang}},
  \bibinfo {author} {\bibfnamefont {H.}~\bibnamefont {Tao}}, \bibinfo {author}
  {\bibfnamefont {A.~C.}\ \bibnamefont {Strikwerda}}, \bibinfo {author}
  {\bibfnamefont {K.}~\bibnamefont {Fan}}, \bibinfo {author} {\bibfnamefont
  {G.~R.}\ \bibnamefont {Keiser}}, \bibinfo {author} {\bibfnamefont {A.~J.}\
  \bibnamefont {Sternbach}}, \bibinfo {author} {\bibfnamefont {K.~G.}\
  \bibnamefont {West}}, \bibinfo {author} {\bibfnamefont {S.}~\bibnamefont
  {Kittiwatanakul}}, \bibinfo {author} {\bibfnamefont {J.}~\bibnamefont {Lu}},
  \bibinfo {author} {\bibfnamefont {S.~A.}\ \bibnamefont {Wolf}}, \bibinfo
  {author} {\bibfnamefont {F.~G.}\ \bibnamefont {Omenetto}}, \bibinfo {author}
  {\bibfnamefont {X.}~\bibnamefont {Zhang}}, \bibinfo {author} {\bibfnamefont
  {K.~A.}\ \bibnamefont {Nelson}}, \ and\ \bibinfo {author} {\bibfnamefont
  {R.~D.}\ \bibnamefont {Averitt}},\ }\href {\doibase 10.1038/nature11231}
  {\bibfield  {journal} {\bibinfo  {journal} {Nature}\ }\textbf {\bibinfo
  {volume} {487}},\ \bibinfo {pages} {345} (\bibinfo {year}
  {2012})}\BibitemShut {NoStop}%
\bibitem [{\citenamefont {Morrison}\ \emph {et~al.}(2014)\citenamefont
  {Morrison}, \citenamefont {Chatelain}, \citenamefont {Tiwari}, \citenamefont
  {Hendaoui}, \citenamefont {Bruh{\'a}cs}, \citenamefont {Chaker},\ and\
  \citenamefont {Siwick}}]{morrison2014}%
  \BibitemOpen
  \bibfield  {author} {\bibinfo {author} {\bibfnamefont {V.~R.}\ \bibnamefont
  {Morrison}}, \bibinfo {author} {\bibfnamefont {{\relax Robert}.~P.}\
  \bibnamefont {Chatelain}}, \bibinfo {author} {\bibfnamefont {K.~L.}\
  \bibnamefont {Tiwari}}, \bibinfo {author} {\bibfnamefont {A.}~\bibnamefont
  {Hendaoui}}, \bibinfo {author} {\bibfnamefont {A.}~\bibnamefont
  {Bruh{\'a}cs}}, \bibinfo {author} {\bibfnamefont {M.}~\bibnamefont {Chaker}},
  \ and\ \bibinfo {author} {\bibfnamefont {B.~J.}\ \bibnamefont {Siwick}},\
  }\href {\doibase 10.1126/science.1253779} {\bibfield  {journal} {\bibinfo
  {journal} {Science}\ }\textbf {\bibinfo {volume} {346}},\ \bibinfo {pages}
  {445} (\bibinfo {year} {2014})}\BibitemShut {NoStop}%
\bibitem [{\citenamefont {Wegkamp}\ \emph {et~al.}(2014)\citenamefont
  {Wegkamp}, \citenamefont {Herzog}, \citenamefont {Xian}, \citenamefont
  {Gatti}, \citenamefont {Cudazzo}, \citenamefont {McGahan}, \citenamefont
  {Marvel}, \citenamefont {Haglund}, \citenamefont {Rubio}, \citenamefont
  {Wolf},\ and\ \citenamefont {St{\"a}hler}}]{wegkamp2014}%
  \BibitemOpen
  \bibfield  {author} {\bibinfo {author} {\bibfnamefont {D.}~\bibnamefont
  {Wegkamp}}, \bibinfo {author} {\bibfnamefont {M.}~\bibnamefont {Herzog}},
  \bibinfo {author} {\bibfnamefont {L.}~\bibnamefont {Xian}}, \bibinfo {author}
  {\bibfnamefont {M.}~\bibnamefont {Gatti}}, \bibinfo {author} {\bibfnamefont
  {P.}~\bibnamefont {Cudazzo}}, \bibinfo {author} {\bibfnamefont {C.~L.}\
  \bibnamefont {McGahan}}, \bibinfo {author} {\bibfnamefont {R.~E.}\
  \bibnamefont {Marvel}}, \bibinfo {author} {\bibfnamefont {R.~F.}\
  \bibnamefont {Haglund}}, \bibinfo {author} {\bibfnamefont {A.}~\bibnamefont
  {Rubio}}, \bibinfo {author} {\bibfnamefont {M.}~\bibnamefont {Wolf}}, \ and\
  \bibinfo {author} {\bibfnamefont {J.}~\bibnamefont {St{\"a}hler}},\ }\href
  {\doibase 10.1103/PhysRevLett.113.216401} {\bibfield  {journal} {\bibinfo
  {journal} {Phys. Rev. Lett.}\ }\textbf {\bibinfo {volume} {113}},\ \bibinfo
  {pages} {216401} (\bibinfo {year} {2014})}\BibitemShut {NoStop}%
\bibitem [{\citenamefont {Laverock}\ \emph {et~al.}(2014)\citenamefont
  {Laverock}, \citenamefont {Kittiwatanakul}, \citenamefont {Zakharov},
  \citenamefont {Niu}, \citenamefont {Chen}, \citenamefont {Wolf},
  \citenamefont {Lu},\ and\ \citenamefont {Smith}}]{laverock2014}%
  \BibitemOpen
  \bibfield  {author} {\bibinfo {author} {\bibfnamefont {J.}~\bibnamefont
  {Laverock}}, \bibinfo {author} {\bibfnamefont {S.}~\bibnamefont
  {Kittiwatanakul}}, \bibinfo {author} {\bibfnamefont {A.~A.}\ \bibnamefont
  {Zakharov}}, \bibinfo {author} {\bibfnamefont {Y.~R.}\ \bibnamefont {Niu}},
  \bibinfo {author} {\bibfnamefont {B.}~\bibnamefont {Chen}}, \bibinfo {author}
  {\bibfnamefont {S.~A.}\ \bibnamefont {Wolf}}, \bibinfo {author}
  {\bibfnamefont {J.~W.}\ \bibnamefont {Lu}}, \ and\ \bibinfo {author}
  {\bibfnamefont {K.~E.}\ \bibnamefont {Smith}},\ }\href {\doibase
  10.1103/PhysRevLett.113.216402} {\bibfield  {journal} {\bibinfo  {journal}
  {Phys. Rev. Lett.}\ }\textbf {\bibinfo {volume} {113}},\ \bibinfo {pages}
  {216402} (\bibinfo {year} {2014})}\BibitemShut {NoStop}%
\bibitem [{\citenamefont {Jager}\ \emph {et~al.}(2017)\citenamefont {Jager},
  \citenamefont {Ott}, \citenamefont {Kraus}, \citenamefont {Kaplan},
  \citenamefont {Pouse}, \citenamefont {Marvel}, \citenamefont {Haglund},
  \citenamefont {Neumark},\ and\ \citenamefont {Leone}}]{jager2017}%
  \BibitemOpen
  \bibfield  {author} {\bibinfo {author} {\bibfnamefont {M.~F.}\ \bibnamefont
  {Jager}}, \bibinfo {author} {\bibfnamefont {C.}~\bibnamefont {Ott}}, \bibinfo
  {author} {\bibfnamefont {P.~M.}\ \bibnamefont {Kraus}}, \bibinfo {author}
  {\bibfnamefont {C.~J.}\ \bibnamefont {Kaplan}}, \bibinfo {author}
  {\bibfnamefont {W.}~\bibnamefont {Pouse}}, \bibinfo {author} {\bibfnamefont
  {R.~E.}\ \bibnamefont {Marvel}}, \bibinfo {author} {\bibfnamefont {R.~F.}\
  \bibnamefont {Haglund}}, \bibinfo {author} {\bibfnamefont {D.~M.}\
  \bibnamefont {Neumark}}, \ and\ \bibinfo {author} {\bibfnamefont {S.~R.}\
  \bibnamefont {Leone}},\ }\href {\doibase 10.1073/pnas.1707602114} {\bibfield
  {journal} {\bibinfo  {journal} {Proceedings of the National Academy of
  Sciences of the United States of America}\ }\textbf {\bibinfo {volume}
  {114}},\ \bibinfo {pages} {9558} (\bibinfo {year} {2017})}\BibitemShut
  {NoStop}%
\bibitem [{\citenamefont {Otto}\ \emph {et~al.}(2019)\citenamefont {Otto},
  \citenamefont {René~de Cotret}, \citenamefont {Valverde-Chavez},
  \citenamefont {Tiwari}, \citenamefont {Émond}, \citenamefont {Chaker},
  \citenamefont {Cooke},\ and\ \citenamefont {Siwick}}]{otto2019}%
  \BibitemOpen
  \bibfield  {author} {\bibinfo {author} {\bibfnamefont {M.~R.}\ \bibnamefont
  {Otto}}, \bibinfo {author} {\bibfnamefont {L.~P.}\ \bibnamefont {René~de
  Cotret}}, \bibinfo {author} {\bibfnamefont {D.~A.}\ \bibnamefont
  {Valverde-Chavez}}, \bibinfo {author} {\bibfnamefont {K.~L.}\ \bibnamefont
  {Tiwari}}, \bibinfo {author} {\bibfnamefont {N.}~\bibnamefont {Émond}},
  \bibinfo {author} {\bibfnamefont {M.}~\bibnamefont {Chaker}}, \bibinfo
  {author} {\bibfnamefont {D.~G.}\ \bibnamefont {Cooke}}, \ and\ \bibinfo
  {author} {\bibfnamefont {B.~J.}\ \bibnamefont {Siwick}},\ }\href {\doibase
  10.1073/pnas.1808414115} {\bibfield  {journal} {\bibinfo  {journal}
  {Proceedings of the National Academy of Sciences}\ }\textbf {\bibinfo
  {volume} {116}},\ \bibinfo {pages} {450} (\bibinfo {year}
  {2019})}\BibitemShut {NoStop}%
\bibitem [{\citenamefont {Vidas}\ \emph {et~al.}(2020)\citenamefont {Vidas},
  \citenamefont {Schick}, \citenamefont {Mart{\'i}nez}, \citenamefont
  {{Perez-Salinas}}, \citenamefont {{Ramos-{\'A}lvarez}}, \citenamefont
  {Cichy}, \citenamefont {{Batlle-Porro}}, \citenamefont {Johnson},
  \citenamefont {Hallman}, \citenamefont {Haglund},\ and\ \citenamefont
  {Wall}}]{vidas2020}%
  \BibitemOpen
  \bibfield  {author} {\bibinfo {author} {\bibfnamefont {L.}~\bibnamefont
  {Vidas}}, \bibinfo {author} {\bibfnamefont {D.}~\bibnamefont {Schick}},
  \bibinfo {author} {\bibfnamefont {E.}~\bibnamefont {Mart{\'i}nez}}, \bibinfo
  {author} {\bibfnamefont {D.}~\bibnamefont {{Perez-Salinas}}}, \bibinfo
  {author} {\bibfnamefont {A.}~\bibnamefont {{Ramos-{\'A}lvarez}}}, \bibinfo
  {author} {\bibfnamefont {S.}~\bibnamefont {Cichy}}, \bibinfo {author}
  {\bibfnamefont {S.}~\bibnamefont {{Batlle-Porro}}}, \bibinfo {author}
  {\bibfnamefont {A.~S.}\ \bibnamefont {Johnson}}, \bibinfo {author}
  {\bibfnamefont {K.~A.}\ \bibnamefont {Hallman}}, \bibinfo {author}
  {\bibfnamefont {R.~F.}\ \bibnamefont {Haglund}}, \ and\ \bibinfo {author}
  {\bibfnamefont {S.}~\bibnamefont {Wall}},\ }\href {\doibase
  10.1103/PhysRevX.10.031047} {\bibfield  {journal} {\bibinfo  {journal} {Phys.
  Rev. X}\ }\textbf {\bibinfo {volume} {10}},\ \bibinfo {pages} {031047}
  (\bibinfo {year} {2020})}\BibitemShut {NoStop}%
\bibitem [{\citenamefont {Xu}\ \emph {et~al.}(2022{\natexlab{a}})\citenamefont
  {Xu}, \citenamefont {Jin}, \citenamefont {Chen}, \citenamefont {Lu},
  \citenamefont {Cheng}, \citenamefont {Zhang}, \citenamefont {Qi},
  \citenamefont {Chen}, \citenamefont {Yin}, \citenamefont {Wang},
  \citenamefont {Xiang},\ and\ \citenamefont {Qian}}]{xu2022v}%
  \BibitemOpen
  \bibfield  {author} {\bibinfo {author} {\bibfnamefont {C.}~\bibnamefont
  {Xu}}, \bibinfo {author} {\bibfnamefont {C.}~\bibnamefont {Jin}}, \bibinfo
  {author} {\bibfnamefont {Z.}~\bibnamefont {Chen}}, \bibinfo {author}
  {\bibfnamefont {Q.}~\bibnamefont {Lu}}, \bibinfo {author} {\bibfnamefont
  {Y.}~\bibnamefont {Cheng}}, \bibinfo {author} {\bibfnamefont
  {B.}~\bibnamefont {Zhang}}, \bibinfo {author} {\bibfnamefont
  {F.}~\bibnamefont {Qi}}, \bibinfo {author} {\bibfnamefont {J.}~\bibnamefont
  {Chen}}, \bibinfo {author} {\bibfnamefont {X.}~\bibnamefont {Yin}}, \bibinfo
  {author} {\bibfnamefont {G.}~\bibnamefont {Wang}}, \bibinfo {author}
  {\bibfnamefont {D.}~\bibnamefont {Xiang}}, \ and\ \bibinfo {author}
  {\bibfnamefont {D.}~\bibnamefont {Qian}},\ }\href {\doibase
  10.48550/arXiv.2203.09776} {\enquote {\bibinfo {title} {{VO2} {Does} {Not}
  {Host} a {Photoinduced} {Long}-lived {Monoclinic} {Metallic} {Phase}},}\ }
  (\bibinfo {year} {2022}{\natexlab{a}})\BibitemShut {NoStop}%
\bibitem [{\citenamefont {Aoki}\ \emph {et~al.}(2014)\citenamefont {Aoki},
  \citenamefont {Tsuji}, \citenamefont {Eckstein}, \citenamefont {Kollar},
  \citenamefont {Oka},\ and\ \citenamefont {Werner}}]{aoki2014}%
  \BibitemOpen
  \bibfield  {author} {\bibinfo {author} {\bibfnamefont {H.}~\bibnamefont
  {Aoki}}, \bibinfo {author} {\bibfnamefont {N.}~\bibnamefont {Tsuji}},
  \bibinfo {author} {\bibfnamefont {M.}~\bibnamefont {Eckstein}}, \bibinfo
  {author} {\bibfnamefont {M.}~\bibnamefont {Kollar}}, \bibinfo {author}
  {\bibfnamefont {T.}~\bibnamefont {Oka}}, \ and\ \bibinfo {author}
  {\bibfnamefont {P.}~\bibnamefont {Werner}},\ }\href {\doibase
  10.1103/RevModPhys.86.779} {\bibfield  {journal} {\bibinfo  {journal} {Rev.
  Mod. Phys.}\ }\textbf {\bibinfo {volume} {86}},\ \bibinfo {pages} {779}
  (\bibinfo {year} {2014})}\BibitemShut {NoStop}%
\bibitem [{\citenamefont {Andersson}\ \emph {et~al.}(1956)\citenamefont
  {Andersson}, \citenamefont {Parck}, \citenamefont {Ulfvarson}, \citenamefont
  {Stenhagen},\ and\ \citenamefont {Thorell}}]{andersson1956}%
  \BibitemOpen
  \bibfield  {author} {\bibinfo {author} {\bibfnamefont {G.}~\bibnamefont
  {Andersson}}, \bibinfo {author} {\bibfnamefont {C.}~\bibnamefont {Parck}},
  \bibinfo {author} {\bibfnamefont {U.}~\bibnamefont {Ulfvarson}}, \bibinfo
  {author} {\bibfnamefont {E.}~\bibnamefont {Stenhagen}}, \ and\ \bibinfo
  {author} {\bibfnamefont {B.}~\bibnamefont {Thorell}},\ }\href {\doibase
  10.3891/acta.chem.scand.10-0623} {\bibfield  {journal} {\bibinfo  {journal}
  {Acta Chemica Scandinavica}\ }\textbf {\bibinfo {volume} {10}},\ \bibinfo
  {pages} {623} (\bibinfo {year} {1956})}\BibitemShut {NoStop}%
\bibitem [{\citenamefont {Longo}\ \emph {et~al.}(1970)\citenamefont {Longo},
  \citenamefont {Kierkegaard}, \citenamefont {Ballhausen}, \citenamefont
  {Ragnarsson}, \citenamefont {Rasmussen}, \citenamefont {Sunde},\ and\
  \citenamefont {S{\o}rensen}}]{longo1970}%
  \BibitemOpen
  \bibfield  {author} {\bibinfo {author} {\bibfnamefont {J.~M.}\ \bibnamefont
  {Longo}}, \bibinfo {author} {\bibfnamefont {P.}~\bibnamefont {Kierkegaard}},
  \bibinfo {author} {\bibfnamefont {C.~J.}\ \bibnamefont {Ballhausen}},
  \bibinfo {author} {\bibfnamefont {U.}~\bibnamefont {Ragnarsson}}, \bibinfo
  {author} {\bibfnamefont {S.~E.}\ \bibnamefont {Rasmussen}}, \bibinfo {author}
  {\bibfnamefont {E.}~\bibnamefont {Sunde}}, \ and\ \bibinfo {author}
  {\bibfnamefont {N.~A.}\ \bibnamefont {S{\o}rensen}},\ }\href {\doibase
  10.3891/acta.chem.scand.24-0420} {\bibfield  {journal} {\bibinfo  {journal}
  {Acta Chemica Scandinavica}\ }\textbf {\bibinfo {volume} {24}},\ \bibinfo
  {pages} {420} (\bibinfo {year} {1970})}\BibitemShut {NoStop}%
\bibitem [{\citenamefont {Giannozzi}\ \emph {et~al.}(2017)\citenamefont
  {Giannozzi}, \citenamefont {Andreussi}, \citenamefont {Brumme}, \citenamefont
  {Bunau}, \citenamefont {Nardelli}, \citenamefont {Calandra}, \citenamefont
  {Car}, \citenamefont {Cavazzoni}, \citenamefont {Ceresoli}, \citenamefont
  {Cococcioni}, \citenamefont {Colonna}, \citenamefont {Carnimeo},
  \citenamefont {Corso}, \citenamefont {de~Gironcoli}, \citenamefont {Delugas},
  \citenamefont {DiStasio}, \citenamefont {Ferretti}, \citenamefont {Floris},
  \citenamefont {Fratesi}, \citenamefont {Fugallo}, \citenamefont {Gebauer},
  \citenamefont {Gerstmann}, \citenamefont {Giustino}, \citenamefont {Gorni},
  \citenamefont {Jia}, \citenamefont {Kawamura}, \citenamefont {Ko},
  \citenamefont {Kokalj}, \citenamefont {K{\"u}{\c c}{\"u}kbenli},
  \citenamefont {Lazzeri}, \citenamefont {Marsili}, \citenamefont {Marzari},
  \citenamefont {Mauri}, \citenamefont {Nguyen}, \citenamefont {Nguyen},
  \citenamefont {{Otero-de-la-Roza}}, \citenamefont {Paulatto}, \citenamefont
  {Ponc{\'e}}, \citenamefont {Rocca}, \citenamefont {Sabatini}, \citenamefont
  {Santra}, \citenamefont {Schlipf}, \citenamefont {Seitsonen}, \citenamefont
  {Smogunov}, \citenamefont {Timrov}, \citenamefont {Thonhauser}, \citenamefont
  {Umari}, \citenamefont {Vast}, \citenamefont {Wu},\ and\ \citenamefont
  {Baroni}}]{giannozzi2017}%
  \BibitemOpen
  \bibfield  {author} {\bibinfo {author} {\bibfnamefont {P.}~\bibnamefont
  {Giannozzi}}, \bibinfo {author} {\bibfnamefont {O.}~\bibnamefont
  {Andreussi}}, \bibinfo {author} {\bibfnamefont {T.}~\bibnamefont {Brumme}},
  \bibinfo {author} {\bibfnamefont {O.}~\bibnamefont {Bunau}}, \bibinfo
  {author} {\bibfnamefont {M.~B.}\ \bibnamefont {Nardelli}}, \bibinfo {author}
  {\bibfnamefont {M.}~\bibnamefont {Calandra}}, \bibinfo {author}
  {\bibfnamefont {R.}~\bibnamefont {Car}}, \bibinfo {author} {\bibfnamefont
  {C.}~\bibnamefont {Cavazzoni}}, \bibinfo {author} {\bibfnamefont
  {D.}~\bibnamefont {Ceresoli}}, \bibinfo {author} {\bibfnamefont
  {M.}~\bibnamefont {Cococcioni}}, \bibinfo {author} {\bibfnamefont
  {N.}~\bibnamefont {Colonna}}, \bibinfo {author} {\bibfnamefont
  {I.}~\bibnamefont {Carnimeo}}, \bibinfo {author} {\bibfnamefont {A.~D.}\
  \bibnamefont {Corso}}, \bibinfo {author} {\bibfnamefont {S.}~\bibnamefont
  {de~Gironcoli}}, \bibinfo {author} {\bibfnamefont {P.}~\bibnamefont
  {Delugas}}, \bibinfo {author} {\bibfnamefont {R.~A.}\ \bibnamefont
  {DiStasio}}, \bibinfo {author} {\bibfnamefont {A.}~\bibnamefont {Ferretti}},
  \bibinfo {author} {\bibfnamefont {A.}~\bibnamefont {Floris}}, \bibinfo
  {author} {\bibfnamefont {G.}~\bibnamefont {Fratesi}}, \bibinfo {author}
  {\bibfnamefont {G.}~\bibnamefont {Fugallo}}, \bibinfo {author} {\bibfnamefont
  {R.}~\bibnamefont {Gebauer}}, \bibinfo {author} {\bibfnamefont
  {U.}~\bibnamefont {Gerstmann}}, \bibinfo {author} {\bibfnamefont
  {F.}~\bibnamefont {Giustino}}, \bibinfo {author} {\bibfnamefont
  {T.}~\bibnamefont {Gorni}}, \bibinfo {author} {\bibfnamefont
  {J.}~\bibnamefont {Jia}}, \bibinfo {author} {\bibfnamefont {M.}~\bibnamefont
  {Kawamura}}, \bibinfo {author} {\bibfnamefont {H.-Y.}\ \bibnamefont {Ko}},
  \bibinfo {author} {\bibfnamefont {A.}~\bibnamefont {Kokalj}}, \bibinfo
  {author} {\bibfnamefont {E.}~\bibnamefont {K{\"u}{\c c}{\"u}kbenli}},
  \bibinfo {author} {\bibfnamefont {M.}~\bibnamefont {Lazzeri}}, \bibinfo
  {author} {\bibfnamefont {M.}~\bibnamefont {Marsili}}, \bibinfo {author}
  {\bibfnamefont {N.}~\bibnamefont {Marzari}}, \bibinfo {author} {\bibfnamefont
  {F.}~\bibnamefont {Mauri}}, \bibinfo {author} {\bibfnamefont {N.~L.}\
  \bibnamefont {Nguyen}}, \bibinfo {author} {\bibfnamefont {H.-V.}\
  \bibnamefont {Nguyen}}, \bibinfo {author} {\bibfnamefont {A.}~\bibnamefont
  {{Otero-de-la-Roza}}}, \bibinfo {author} {\bibfnamefont {L.}~\bibnamefont
  {Paulatto}}, \bibinfo {author} {\bibfnamefont {S.}~\bibnamefont {Ponc{\'e}}},
  \bibinfo {author} {\bibfnamefont {D.}~\bibnamefont {Rocca}}, \bibinfo
  {author} {\bibfnamefont {R.}~\bibnamefont {Sabatini}}, \bibinfo {author}
  {\bibfnamefont {B.}~\bibnamefont {Santra}}, \bibinfo {author} {\bibfnamefont
  {M.}~\bibnamefont {Schlipf}}, \bibinfo {author} {\bibfnamefont {A.~P.}\
  \bibnamefont {Seitsonen}}, \bibinfo {author} {\bibfnamefont {A.}~\bibnamefont
  {Smogunov}}, \bibinfo {author} {\bibfnamefont {I.}~\bibnamefont {Timrov}},
  \bibinfo {author} {\bibfnamefont {T.}~\bibnamefont {Thonhauser}}, \bibinfo
  {author} {\bibfnamefont {P.}~\bibnamefont {Umari}}, \bibinfo {author}
  {\bibfnamefont {N.}~\bibnamefont {Vast}}, \bibinfo {author} {\bibfnamefont
  {X.}~\bibnamefont {Wu}}, \ and\ \bibinfo {author} {\bibfnamefont
  {S.}~\bibnamefont {Baroni}},\ }\href {\doibase 10.1088/1361-648X/aa8f79}
  {\bibfield  {journal} {\bibinfo  {journal} {J. Phys.: Condens. Matter}\
  }\textbf {\bibinfo {volume} {29}},\ \bibinfo {pages} {465901} (\bibinfo
  {year} {2017})}\BibitemShut {NoStop}%
\bibitem [{\citenamefont {Pizzi}\ \emph {et~al.}(2020)\citenamefont {Pizzi},
  \citenamefont {Vitale}, \citenamefont {Arita}, \citenamefont {Bl{\"u}gel},
  \citenamefont {Freimuth}, \citenamefont {G{\'e}ranton}, \citenamefont
  {Gibertini}, \citenamefont {Gresch}, \citenamefont {Johnson}, \citenamefont
  {Koretsune}, \citenamefont {{Iba{\~n}ez-Azpiroz}}, \citenamefont {Lee},
  \citenamefont {Lihm}, \citenamefont {Marchand}, \citenamefont {Marrazzo},
  \citenamefont {Mokrousov}, \citenamefont {Mustafa}, \citenamefont {Nohara},
  \citenamefont {Nomura}, \citenamefont {Paulatto}, \citenamefont {Ponc{\'e}},
  \citenamefont {Ponweiser}, \citenamefont {Qiao}, \citenamefont {Th{\"o}le},
  \citenamefont {Tsirkin}, \citenamefont {Wierzbowska}, \citenamefont
  {Marzari}, \citenamefont {Vanderbilt}, \citenamefont {Souza}, \citenamefont
  {Mostofi},\ and\ \citenamefont {Yates}}]{pizzi2020}%
  \BibitemOpen
  \bibfield  {author} {\bibinfo {author} {\bibfnamefont {G.}~\bibnamefont
  {Pizzi}}, \bibinfo {author} {\bibfnamefont {V.}~\bibnamefont {Vitale}},
  \bibinfo {author} {\bibfnamefont {R.}~\bibnamefont {Arita}}, \bibinfo
  {author} {\bibfnamefont {S.}~\bibnamefont {Bl{\"u}gel}}, \bibinfo {author}
  {\bibfnamefont {F.}~\bibnamefont {Freimuth}}, \bibinfo {author}
  {\bibfnamefont {G.}~\bibnamefont {G{\'e}ranton}}, \bibinfo {author}
  {\bibfnamefont {M.}~\bibnamefont {Gibertini}}, \bibinfo {author}
  {\bibfnamefont {D.}~\bibnamefont {Gresch}}, \bibinfo {author} {\bibfnamefont
  {C.}~\bibnamefont {Johnson}}, \bibinfo {author} {\bibfnamefont
  {T.}~\bibnamefont {Koretsune}}, \bibinfo {author} {\bibfnamefont
  {J.}~\bibnamefont {{Iba{\~n}ez-Azpiroz}}}, \bibinfo {author} {\bibfnamefont
  {H.}~\bibnamefont {Lee}}, \bibinfo {author} {\bibfnamefont {J.-M.}\
  \bibnamefont {Lihm}}, \bibinfo {author} {\bibfnamefont {D.}~\bibnamefont
  {Marchand}}, \bibinfo {author} {\bibfnamefont {A.}~\bibnamefont {Marrazzo}},
  \bibinfo {author} {\bibfnamefont {Y.}~\bibnamefont {Mokrousov}}, \bibinfo
  {author} {\bibfnamefont {J.~I.}\ \bibnamefont {Mustafa}}, \bibinfo {author}
  {\bibfnamefont {Y.}~\bibnamefont {Nohara}}, \bibinfo {author} {\bibfnamefont
  {Y.}~\bibnamefont {Nomura}}, \bibinfo {author} {\bibfnamefont
  {L.}~\bibnamefont {Paulatto}}, \bibinfo {author} {\bibfnamefont
  {S.}~\bibnamefont {Ponc{\'e}}}, \bibinfo {author} {\bibfnamefont
  {T.}~\bibnamefont {Ponweiser}}, \bibinfo {author} {\bibfnamefont
  {J.}~\bibnamefont {Qiao}}, \bibinfo {author} {\bibfnamefont {F.}~\bibnamefont
  {Th{\"o}le}}, \bibinfo {author} {\bibfnamefont {S.~S.}\ \bibnamefont
  {Tsirkin}}, \bibinfo {author} {\bibfnamefont {M.}~\bibnamefont
  {Wierzbowska}}, \bibinfo {author} {\bibfnamefont {N.}~\bibnamefont
  {Marzari}}, \bibinfo {author} {\bibfnamefont {D.}~\bibnamefont {Vanderbilt}},
  \bibinfo {author} {\bibfnamefont {I.}~\bibnamefont {Souza}}, \bibinfo
  {author} {\bibfnamefont {A.~A.}\ \bibnamefont {Mostofi}}, \ and\ \bibinfo
  {author} {\bibfnamefont {J.~R.}\ \bibnamefont {Yates}},\ }\href {\doibase
  10.1088/1361-648X/ab51ff} {\bibfield  {journal} {\bibinfo  {journal} {J.
  Phys.: Condens. Matter}\ }\textbf {\bibinfo {volume} {32}},\ \bibinfo {pages}
  {165902} (\bibinfo {year} {2020})}\BibitemShut {NoStop}%
\bibitem [{\citenamefont {Aryasetiawan}\ \emph {et~al.}(2004)\citenamefont
  {Aryasetiawan}, \citenamefont {Imada}, \citenamefont {Georges}, \citenamefont
  {Kotliar}, \citenamefont {Biermann},\ and\ \citenamefont
  {Lichtenstein}}]{aryasetiawan2004}%
  \BibitemOpen
  \bibfield  {author} {\bibinfo {author} {\bibfnamefont {F.}~\bibnamefont
  {Aryasetiawan}}, \bibinfo {author} {\bibfnamefont {M.}~\bibnamefont {Imada}},
  \bibinfo {author} {\bibfnamefont {A.}~\bibnamefont {Georges}}, \bibinfo
  {author} {\bibfnamefont {G.}~\bibnamefont {Kotliar}}, \bibinfo {author}
  {\bibfnamefont {S.}~\bibnamefont {Biermann}}, \ and\ \bibinfo {author}
  {\bibfnamefont {A.~I.}\ \bibnamefont {Lichtenstein}},\ }\href {\doibase
  10.1103/PhysRevB.70.195104} {\bibfield  {journal} {\bibinfo  {journal} {Phys.
  Rev. B}\ }\textbf {\bibinfo {volume} {70}},\ \bibinfo {pages} {195104}
  (\bibinfo {year} {2004})}\BibitemShut {NoStop}%
\bibitem [{\citenamefont {Nakamura}\ \emph {et~al.}(2021)\citenamefont
  {Nakamura}, \citenamefont {Yoshimoto}, \citenamefont {Nomura}, \citenamefont
  {Tadano}, \citenamefont {Kawamura}, \citenamefont {Kosugi}, \citenamefont
  {Yoshimi}, \citenamefont {Misawa},\ and\ \citenamefont
  {Motoyama}}]{nakamura2021}%
  \BibitemOpen
  \bibfield  {author} {\bibinfo {author} {\bibfnamefont {K.}~\bibnamefont
  {Nakamura}}, \bibinfo {author} {\bibfnamefont {Y.}~\bibnamefont {Yoshimoto}},
  \bibinfo {author} {\bibfnamefont {Y.}~\bibnamefont {Nomura}}, \bibinfo
  {author} {\bibfnamefont {T.}~\bibnamefont {Tadano}}, \bibinfo {author}
  {\bibfnamefont {M.}~\bibnamefont {Kawamura}}, \bibinfo {author}
  {\bibfnamefont {T.}~\bibnamefont {Kosugi}}, \bibinfo {author} {\bibfnamefont
  {K.}~\bibnamefont {Yoshimi}}, \bibinfo {author} {\bibfnamefont
  {T.}~\bibnamefont {Misawa}}, \ and\ \bibinfo {author} {\bibfnamefont
  {Y.}~\bibnamefont {Motoyama}},\ }\href {\doibase 10.1016/j.cpc.2020.107781}
  {\bibfield  {journal} {\bibinfo  {journal} {Computer Physics Communications}\
  }\textbf {\bibinfo {volume} {261}},\ \bibinfo {pages} {107781} (\bibinfo
  {year} {2021})}\BibitemShut {NoStop}%
\bibitem [{\citenamefont {Eckstein}\ and\ \citenamefont
  {Werner}(2013)}]{eckstein2013}%
  \BibitemOpen
  \bibfield  {author} {\bibinfo {author} {\bibfnamefont {M.}~\bibnamefont
  {Eckstein}}\ and\ \bibinfo {author} {\bibfnamefont {P.}~\bibnamefont
  {Werner}},\ }\href {\doibase 10.1103/PhysRevB.88.075135} {\bibfield
  {journal} {\bibinfo  {journal} {Phys. Rev. B}\ }\textbf {\bibinfo {volume}
  {88}},\ \bibinfo {pages} {075135} (\bibinfo {year} {2013})}\BibitemShut
  {NoStop}%
\bibitem [{\citenamefont {Petocchi}\ \emph {et~al.}(2023)\citenamefont
  {Petocchi}, \citenamefont {Chen}, \citenamefont {Li}, \citenamefont
  {Eckstein},\ and\ \citenamefont {Werner}}]{petocchi2023}%
  \BibitemOpen
  \bibfield  {author} {\bibinfo {author} {\bibfnamefont {F.}~\bibnamefont
  {Petocchi}}, \bibinfo {author} {\bibfnamefont {J.}~\bibnamefont {Chen}},
  \bibinfo {author} {\bibfnamefont {J.}~\bibnamefont {Li}}, \bibinfo {author}
  {\bibfnamefont {M.}~\bibnamefont {Eckstein}}, \ and\ \bibinfo {author}
  {\bibfnamefont {P.}~\bibnamefont {Werner}},\ }\href {\doibase
  10.1103/PhysRevB.107.165102} {\bibfield  {journal} {\bibinfo  {journal}
  {Phys. Rev. B}\ }\textbf {\bibinfo {volume} {107}},\ \bibinfo {pages}
  {165102} (\bibinfo {year} {2023})}\BibitemShut {NoStop}%
\bibitem [{\citenamefont {Keiter}\ and\ \citenamefont
  {Kimball}(1971)}]{keiter1971}%
  \BibitemOpen
  \bibfield  {author} {\bibinfo {author} {\bibfnamefont {H.}~\bibnamefont
  {Keiter}}\ and\ \bibinfo {author} {\bibfnamefont {J.~C.}\ \bibnamefont
  {Kimball}},\ }\href {\doibase 10.1063/1.1660293} {\bibfield  {journal}
  {\bibinfo  {journal} {Journal of Applied Physics}\ }\textbf {\bibinfo
  {volume} {42}},\ \bibinfo {pages} {1460} (\bibinfo {year}
  {1971})}\BibitemShut {NoStop}%
\bibitem [{\citenamefont {Eckstein}\ and\ \citenamefont
  {Werner}(2010)}]{eckstein2010}%
  \BibitemOpen
  \bibfield  {author} {\bibinfo {author} {\bibfnamefont {M.}~\bibnamefont
  {Eckstein}}\ and\ \bibinfo {author} {\bibfnamefont {P.}~\bibnamefont
  {Werner}},\ }\href {\doibase 10.1103/PhysRevB.82.115115} {\bibfield
  {journal} {\bibinfo  {journal} {Phys. Rev. B}\ }\textbf {\bibinfo {volume}
  {82}},\ \bibinfo {pages} {115115} (\bibinfo {year} {2010})}\BibitemShut
  {NoStop}%
\bibitem [{\citenamefont {Petocchi}\ \emph {et~al.}(2022)\citenamefont
  {Petocchi}, \citenamefont {Nicholson}, \citenamefont {Salzmann},
  \citenamefont {Pasquier}, \citenamefont {Yazyev}, \citenamefont {Monney},\
  and\ \citenamefont {Werner}}]{petocchi2022}%
  \BibitemOpen
  \bibfield  {author} {\bibinfo {author} {\bibfnamefont {F.}~\bibnamefont
  {Petocchi}}, \bibinfo {author} {\bibfnamefont {C.~W.}\ \bibnamefont
  {Nicholson}}, \bibinfo {author} {\bibfnamefont {B.}~\bibnamefont {Salzmann}},
  \bibinfo {author} {\bibfnamefont {D.}~\bibnamefont {Pasquier}}, \bibinfo
  {author} {\bibfnamefont {O.~V.}\ \bibnamefont {Yazyev}}, \bibinfo {author}
  {\bibfnamefont {C.}~\bibnamefont {Monney}}, \ and\ \bibinfo {author}
  {\bibfnamefont {P.}~\bibnamefont {Werner}},\ }\href {\doibase
  10.1103/PhysRevLett.129.016402} {\bibfield  {journal} {\bibinfo  {journal}
  {Phys. Rev. Lett.}\ }\textbf {\bibinfo {volume} {129}},\ \bibinfo {pages}
  {016402} (\bibinfo {year} {2022})}\BibitemShut {NoStop}%
\bibitem [{\citenamefont {He}\ and\ \citenamefont {Millis}(2016)}]{he2016}%
  \BibitemOpen
  \bibfield  {author} {\bibinfo {author} {\bibfnamefont {Z.}~\bibnamefont
  {He}}\ and\ \bibinfo {author} {\bibfnamefont {A.~J.}\ \bibnamefont
  {Millis}},\ }\href {\doibase 10.1103/PhysRevB.93.115126} {\bibfield
  {journal} {\bibinfo  {journal} {Phys. Rev. B}\ }\textbf {\bibinfo {volume}
  {93}},\ \bibinfo {pages} {115126} (\bibinfo {year} {2016})}\BibitemShut
  {NoStop}%
\bibitem [{\citenamefont {Xu}\ \emph {et~al.}(2022{\natexlab{b}})\citenamefont
  {Xu}, \citenamefont {Chen},\ and\ \citenamefont {Meng}}]{xu2022}%
  \BibitemOpen
  \bibfield  {author} {\bibinfo {author} {\bibfnamefont {J.}~\bibnamefont
  {Xu}}, \bibinfo {author} {\bibfnamefont {D.}~\bibnamefont {Chen}}, \ and\
  \bibinfo {author} {\bibfnamefont {S.}~\bibnamefont {Meng}},\ }\href {\doibase
  10.1126/sciadv.add2392} {\bibfield  {journal} {\bibinfo  {journal} {Sci.
  Adv.}\ }\textbf {\bibinfo {volume} {8}},\ \bibinfo {pages} {eadd2392}
  (\bibinfo {year} {2022}{\natexlab{b}})}\BibitemShut {NoStop}%
\bibitem [{\citenamefont {Schüler}\ \emph {et~al.}(2020)\citenamefont
  {Schüler}, \citenamefont {Golež}, \citenamefont {Murakami}, \citenamefont
  {Bittner}, \citenamefont {Herrmann}, \citenamefont {Strand}, \citenamefont
  {Werner},\ and\ \citenamefont {Eckstein}}]{schuler2020}%
  \BibitemOpen
  \bibfield  {author} {\bibinfo {author} {\bibfnamefont {M.}~\bibnamefont
  {Schüler}}, \bibinfo {author} {\bibfnamefont {D.}~\bibnamefont {Golež}},
  \bibinfo {author} {\bibfnamefont {Y.}~\bibnamefont {Murakami}}, \bibinfo
  {author} {\bibfnamefont {N.}~\bibnamefont {Bittner}}, \bibinfo {author}
  {\bibfnamefont {A.}~\bibnamefont {Herrmann}}, \bibinfo {author}
  {\bibfnamefont {H.~U.~R.}\ \bibnamefont {Strand}}, \bibinfo {author}
  {\bibfnamefont {P.}~\bibnamefont {Werner}}, \ and\ \bibinfo {author}
  {\bibfnamefont {M.}~\bibnamefont {Eckstein}},\ }\href {\doibase
  10.1016/j.cpc.2020.107484} {\bibfield  {journal} {\bibinfo  {journal}
  {Computer Physics Communications}\ }\textbf {\bibinfo {volume} {257}},\
  \bibinfo {pages} {107484} (\bibinfo {year} {2020})}\BibitemShut {NoStop}%
\bibitem [{\citenamefont {Lichtenstein}\ and\ \citenamefont
  {Katsnelson}(2000)}]{lichtenstein2000}%
  \BibitemOpen
  \bibfield  {author} {\bibinfo {author} {\bibfnamefont {A.~I.}\ \bibnamefont
  {Lichtenstein}}\ and\ \bibinfo {author} {\bibfnamefont {M.~I.}\ \bibnamefont
  {Katsnelson}},\ }\href {\doibase 10.1103/PhysRevB.62.R9283} {\bibfield
  {journal} {\bibinfo  {journal} {Phys. Rev. B}\ }\textbf {\bibinfo {volume}
  {62}},\ \bibinfo {pages} {R9283} (\bibinfo {year} {2000})}\BibitemShut
  {NoStop}%
\bibitem [{\citenamefont {Werner}\ \emph {et~al.}(2017)\citenamefont {Werner},
  \citenamefont {Strand}, \citenamefont {Hoshino},\ and\ \citenamefont
  {Eckstein}}]{werner2017}%
  \BibitemOpen
  \bibfield  {author} {\bibinfo {author} {\bibfnamefont {P.}~\bibnamefont
  {Werner}}, \bibinfo {author} {\bibfnamefont {H.~U.~R.}\ \bibnamefont
  {Strand}}, \bibinfo {author} {\bibfnamefont {S.}~\bibnamefont {Hoshino}}, \
  and\ \bibinfo {author} {\bibfnamefont {M.}~\bibnamefont {Eckstein}},\ }\href
  {\doibase 10.1103/PhysRevB.95.195405} {\bibfield  {journal} {\bibinfo
  {journal} {Phys. Rev. B}\ }\textbf {\bibinfo {volume} {95}},\ \bibinfo
  {pages} {195405} (\bibinfo {year} {2017})}\BibitemShut {NoStop}%
\bibitem [{\citenamefont {Petocchi}\ \emph {et~al.}(2019)\citenamefont
  {Petocchi}, \citenamefont {Beck}, \citenamefont {Ederer},\ and\ \citenamefont
  {Werner}}]{petocchi2019}%
  \BibitemOpen
  \bibfield  {author} {\bibinfo {author} {\bibfnamefont {F.}~\bibnamefont
  {Petocchi}}, \bibinfo {author} {\bibfnamefont {S.}~\bibnamefont {Beck}},
  \bibinfo {author} {\bibfnamefont {C.}~\bibnamefont {Ederer}}, \ and\ \bibinfo
  {author} {\bibfnamefont {P.}~\bibnamefont {Werner}},\ }\href {\doibase
  10.1103/PhysRevB.100.075147} {\bibfield  {journal} {\bibinfo  {journal}
  {Phys. Rev. B}\ }\textbf {\bibinfo {volume} {100}},\ \bibinfo {pages}
  {075147} (\bibinfo {year} {2019})}\BibitemShut {NoStop}%
\bibitem [{\citenamefont {Georges}\ \emph {et~al.}(1996)\citenamefont
  {Georges}, \citenamefont {Kotliar}, \citenamefont {Krauth},\ and\
  \citenamefont {Rozenberg}}]{georges1996}%
  \BibitemOpen
  \bibfield  {author} {\bibinfo {author} {\bibfnamefont {A.}~\bibnamefont
  {Georges}}, \bibinfo {author} {\bibfnamefont {G.}~\bibnamefont {Kotliar}},
  \bibinfo {author} {\bibfnamefont {W.}~\bibnamefont {Krauth}}, \ and\ \bibinfo
  {author} {\bibfnamefont {M.~J.}\ \bibnamefont {Rozenberg}},\ }\href {\doibase
  10.1103/RevModPhys.68.13} {\bibfield  {journal} {\bibinfo  {journal} {Rev.
  Mod. Phys.}\ }\textbf {\bibinfo {volume} {68}},\ \bibinfo {pages} {13}
  (\bibinfo {year} {1996})}\BibitemShut {NoStop}%
\bibitem [{\citenamefont {de' Medici}\ \emph {et~al.}(2011)\citenamefont {de'
  Medici}, \citenamefont {Mravlje},\ and\ \citenamefont
  {Georges}}]{de_medici_janus-faced_2011}%
  \BibitemOpen
  \bibfield  {author} {\bibinfo {author} {\bibfnamefont {L.}~\bibnamefont {de'
  Medici}}, \bibinfo {author} {\bibfnamefont {J.}~\bibnamefont {Mravlje}}, \
  and\ \bibinfo {author} {\bibfnamefont {A.}~\bibnamefont {Georges}},\ }\href
  {\doibase 10.1103/PhysRevLett.107.256401} {\bibfield  {journal} {\bibinfo
  {journal} {Phys. Rev. Lett.}\ }\textbf {\bibinfo {volume} {107}},\ \bibinfo
  {pages} {256401} (\bibinfo {year} {2011})}\BibitemShut {NoStop}%
\end{thebibliography}%
   
\onecolumngrid   
\appendix

\clearpage

\begin{center}
{\Large \bf Nature of the photo-induced metallic state in monoclinic VO$_2$ \\ Supplemental Material}
\end{center}

\section{Nonequilibrium cDMFT}  

To study the interacting system with electric field excitation, we use the nonequilibrium generalization of cDMFT \cite{lichtenstein2000,aoki2014} with the non-crossing approximation (NCA) \cite{keiter1971,eckstein2010} as cluster impurity solver. Given the complexity of the system, with four V atoms (two dimers) in one unit cell and three t$_{2g}$ orbitals per V atom, we use a simplified self-consistency which is adequate for strongly correlated systems. This approach is based on a Bethe-lattice-inspired real-space construction of the impurity hybridization function. It loses the information on the details of the energy dispersion, but is very economical in terms of memory requirement, which is helpful for nonequilibrium applications~\cite{werner2017,petocchi2019,petocchi2023}.

The time-dependent hybridization function for impurity cluster $i$ can be written as
\begin{equation}\label{eq:hyb}
	\hat{\Delta}_{i}\left(t,t'\right)=\underset{j\neq i}{\sum}\hat{h}_{ij}\left(t\right)\hat{G}_{j}^{\left[i\right]}\left(t,t'\right)\hat{h}_{ji}\left(t'\right),
\end{equation}
where $\hat{h}_{ij}\left(t\right)$ is the time-dependent hopping matrix between the clusters $i$ and $j$, and $\hat{G}_{j}^{\left[i\right]}$ is the cavity Green's function for the lattice with the $i$-th cluster removed \cite{georges1996}. The internal indices (site, orbital, spin) are not explicitly shown, but taken into account by the matrix structure. By approximating the cavity Green's function $\hat{G}_{j}^{\left[i\right]}$ with the (cluster) Green's function  $\hat{G}_{j}$ -- which is only exact in a system with infinite coordination number -- one obtains a self-consistency condition relating the hybridization function directly to the (cluster) Green's function \cite{werner2017,petocchi2019,petocchi2023}, similar to the case of the infinitely connected Bethe lattice \cite{georges1996}. We choose impurity clusters, rather than single impurity atoms, in order to capture the strong nonlocal correlations within the V-V dimers.

The unit cell of M1 VO$_2$ contains two V-V dimers $a$ and $b$, and for each dimer, we define a four-orbital cluster containing the $d_{x^2-y^2}$ and $d_{xz}$ orbitals (impurity cluster $1$), and a two-orbital cluster containing the $d_{yz}$ orbitals (impurity cluster $2$). These clusters allow us to treat the strong intra-dimer hopping within the impurity model. We retain the Slater-Kanamori type interaction within the four-orbital clusters, and the local Hubbard interactions within the two-orbital clusters, whereas the interactions between the $d_{x^2-y^2,xz}$ and $d_{yz}$ orbitals on a given site are treated at the Hartree level. Previous experimental and theoretical studies have shown that the $d_{yz}$ orbital is less relevant for the optical excitations \cite{morrison2014} and also our ED analysis suggests that the $d_{yz}$ orbital is less relevant because of the higher local energy.

In the cDMFT simulations, we hence treat four impurity clusters, with indices $a1$, $a2$, $b1$, $b2$ and $4$, $2$, $4$, $2$ orbitals, respectively. Associated with each cluster impurity model is a $4\times 4$ (or $2\times 2$) contour hybridization function $\hat\Delta\left(t,t'\right)$ constructed from the DFT-derived hopping amplitudes and the $4\times 4$ (or $2\times 2$) contour Green's functions $\hat G(t,t')$. Once all the cluster Green's functions have been obtained by the NCA solver, the hybridization function can be updated using Eq.~(\ref{eq:hyb}) and used as input for the next iteration. Since there are on average two electrons in one V-V dimer, we only keep the Hilbert space sectors with up to three electrons in the NCA solver. This implementation is appropriate for the M1 phase of VO$_2$, where the intra-cluster hopping is larger than the inter-cluster hoppings. 

The Green's function defines the time-dependent density matrix as $\hat{\rho}(t) = i\hat G^{\text{les}}(t,t)$. For any local operator $\hat{O}$, we can then calculate the corresponding expectation value $O(t)=\text{Tr}(\hat{\rho}(t)\hat{O})$. In particular, for a state $|\psi\rangle$, we can define the density matrix (projection operator) $\hat{\rho}_\psi=|\psi\rangle\langle\psi|$ and measure the fidelity $F(t)=\text{Tr}( \hat\rho(t) \hat{\rho}_\psi)$. The electron number operator for a given orbital $\alpha$ is $\hat{n}_\alpha=\hat{c}^\dagger_\alpha \hat{c}_\alpha$, so that $n_\alpha(t)=\text{Tr}(\hat{\rho}(t)\hat{n}_\alpha )$.

\section{Exact diagonalization analysis of the V-V dimer} \label{sec:ED}

To analyze the response of the system to optical excitations, we use exact diagonalization (ED) to solve a time-dependent Kanamori-Hubbard dimer, with Hamiltonian
\begin{eqnarray}
	{H}(t) =H_{\text{tb}}(t)+\sum_i{H_{\text{K}}}_{i}-\mu\sum_{il \sigma} n_{i l \sigma},
\end{eqnarray}
where the hopping Hamiltonian $H_{\text{tb}}(t=0)$ is determined by the DFT calculation,
\begin{eqnarray}
	{H}_{\text{tb}}(t=0) =\sum_{ij} \sum_{lm,\sigma} \left[h^{ij}_{lm} d_{i l \sigma}^{\dagger} d_{j m \sigma}+h.c.\right],
\end{eqnarray}
and the laser pulse is incorporated through the time-dependent Peierls phase $\phi_{ij}(t) = \vec{A}(t)\cdot\vec{r}_{ij}$ 
of the hopping parameters: $h^{ij}_{lm}(t) = h^{ij}_{lm} \text{e}^{-i\frac{e}{\hbar}\phi_{ij}(t)}$. The Kanamori interaction on site $i$ is
	\begin{eqnarray}
		{H_{\text{K}}}_{i} = \sum_{l} U_l n_{il \uparrow} n_{il \downarrow} + \sum_{l \neq m} U'_{lm} n_{il \uparrow} n_{im \downarrow}+ \sum_{l<m, \sigma} (U'_{lm}-J)n_{il \sigma} n_{im \sigma}-J \sum_{ l \neq m} d_{il \uparrow}^{\dagger} d_{il \downarrow}d_{im \downarrow}^{\dagger} d_{im \uparrow}+J \sum_{l \neq m} d_{il \uparrow}^{\dagger} d^{\dagger}_{il \downarrow}d_{im \downarrow} d_{im \uparrow}.\hspace{5mm}
	\end{eqnarray}
Here, $i,j \in \{1,2\}$ are the indices of the two V atoms in the dimer, $\sigma\in \{\uparrow,\downarrow\}$ denotes the electron spin, and $l,m\in \{1,2,3\}$ are the indices of the three $t_{2g}$ orbitals. The interactions $U$, $U'$ and the Hund's coupling $J$ are the cRPA values mentioned in the main text. The form of the electric field pulse $\vec{E}(t)$, which determines the Peierls factor via $\vec{E}(t)=-\partial_t \vec{A}(t)$, is the same as defined in the main text. 
	
\begin{figure}[ht]
	\centering
	\includegraphics[width=0.49\linewidth]{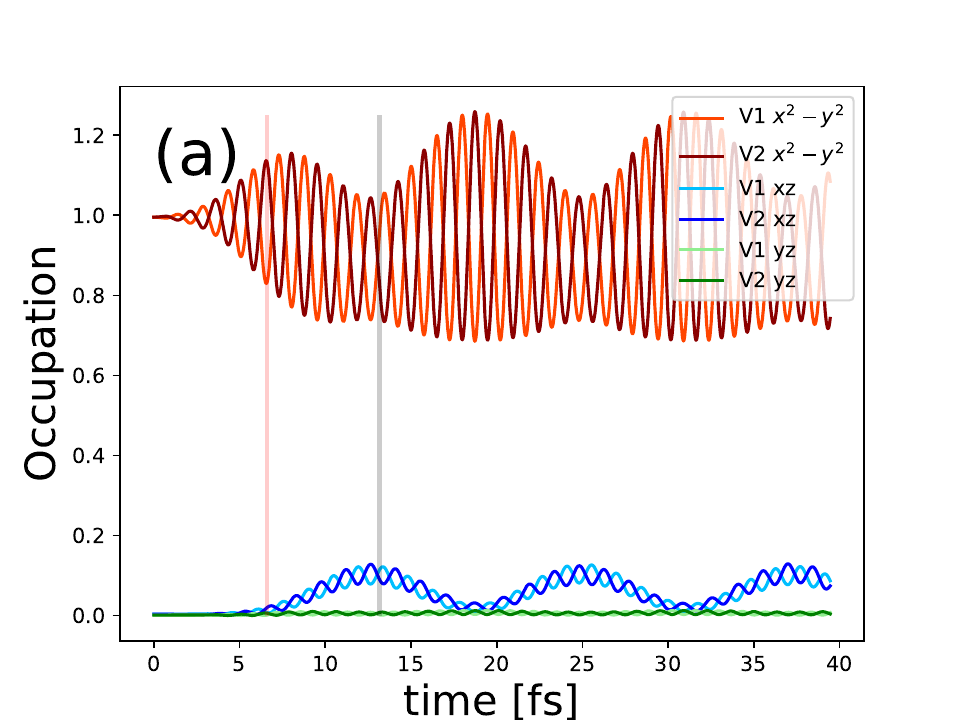}
	\includegraphics[width=0.49\linewidth]{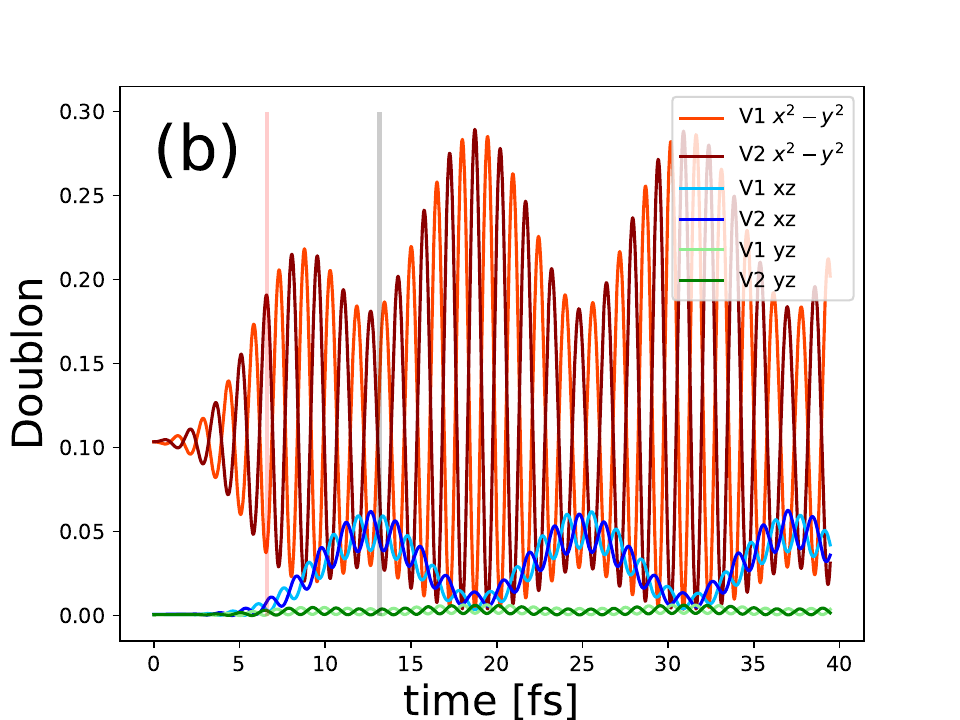}
	\caption{ ED calculation of the occupation (a) and double occupation (b) of each orbital of a V-V dimer. The red and gray vertical lines indicate the maximum and end of the laser pulse. A pulse with frequency $\omega_0=2.8$~eV is used.}
	\label{fig:evolution}
\end{figure}

Figure~\ref{fig:evolution} presents simulation results similar to those in Fig.~4(a,c) of the main text. Panel (a) shows that  the pulse triggers charge oscillation between the two dimer sites and transfers electrons from the $d_{x^2-y^2}$ to the $d_{xz}$ orbital. In panel (b), which shows the evolution of the double occupation, direct evidence for the creation of a  {\it mixed-orbital doublon state} is presented. In these panels, the vertical red and gray lines indicate the maximum and end of the pulse, respectively. Since in the ED analysis, the single dimer is isolated, the energy spectrum is discrete and the time evolution is periodic, without damping.
	
\begin{figure}[ht]
	\centering
	\includegraphics[width=1\linewidth]{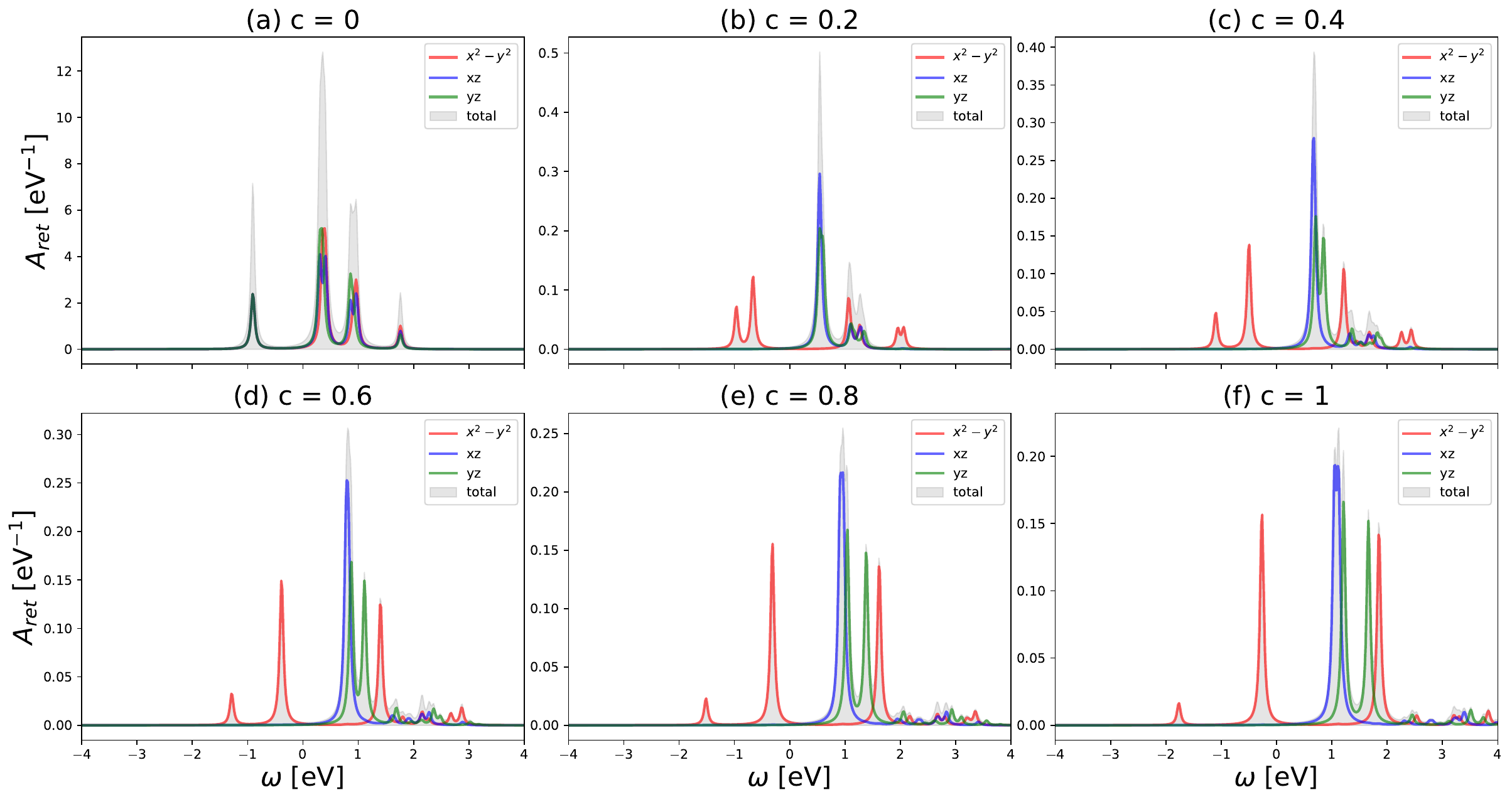}
	\caption{ED equilibrium spectra of the dimer. From (a) to (f), the scaling factor $c$ of the hopping term in the Hamiltonian is increased from 0 (atomic limit) to 1 (realistic value). }
	\label{fig:spectra_ED_hopping}
\end{figure}

\begin{figure}[ht]
	\centering
	\includegraphics[width=\linewidth]{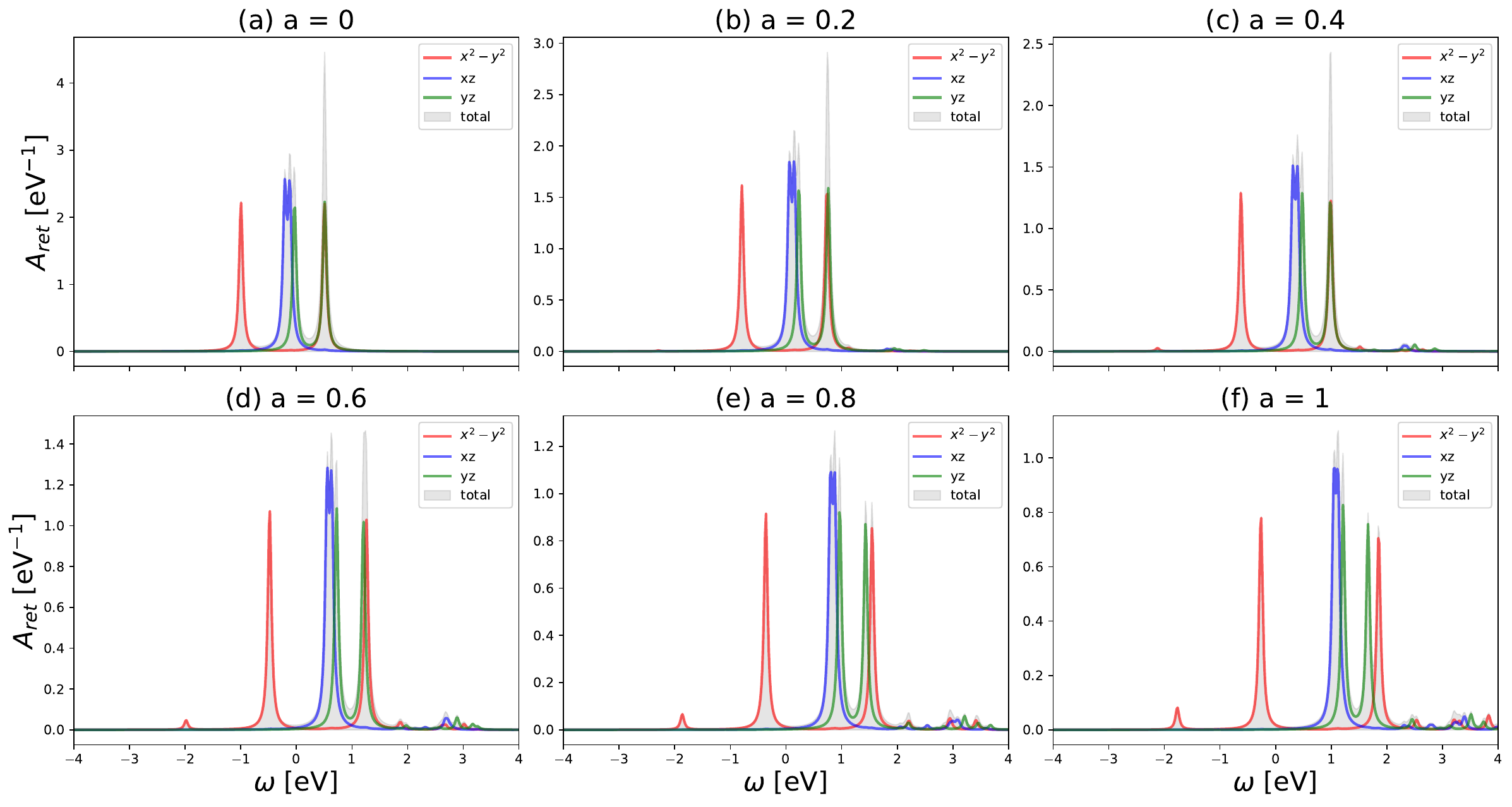}
	\caption{ED equilibrium spectra of the dimer. From (a) to (f), the scaling factor $a$ of the Kanamori interaction in the Hamiltonian is increased from 0 (non-interacting limit) to 1 (realistic value). }
	\label{fig:spectra_ED_interaction}
\end{figure}

In Fig.~\ref{fig:spectra_ED_hopping} we show the equilibrium ED spectra for rescaled hopping amplitudes (rescaling factor $c$) and in Fig.~\ref{fig:spectra_ED_interaction} those for rescaled interactions (rescaling factor $a$), i.e., for the Hamiltonian  
\begin{eqnarray}
	{H} =cH_{\text{tb}}+a\sum_i{H_{\text{K}}}_{i}-\mu\sum_{il \sigma} n_{i l \sigma}.
\end{eqnarray}
Figure~\ref{fig:spectra_ED_hopping}(a) with $c=0$ corresponds to the atomic limit. Here, to better visualize the poles $p$ with weight $A_p$ obtained from the Lehmann representation, we plot $A(\omega)=\sum_p\eta A_p/(\omega-\omega_p+\eta^2)$, with broadening $\eta=0.04$. The upper Hubbard band splits into three subpeaks, corresponding to the creation of different doublon states. As explained for example in Ref.~\onlinecite{de_medici_janus-faced_2011}, in the atomic limit, the three-orbital Kanamori interaction leads to gaps with size  $U-3J$, $U-J$, $U+2J$, corresponding to $1.2$, $1.7$, $2.6$ eV for our parameter set. As the intradimer hopping is turned on ($c>0$), the lower Hubbard band is split into bonding-antibonding peaks, see panels (b)-(f). Figure~\ref{fig:spectra_ED_hopping}(f) with $c=1$ shows the ED spectra for the actual model parameters. The gap around the Fermi energy is formed by peaks of $d_{x^2-y^2}$ and $d_{xz}$ character and corresponds to the inter-orbital same-spin Hubbard interaction $U-3J$. 
	
Figure~\ref{fig:spectra_ED_interaction}(a) with $a=0$ shows the ED spectrum in the non-interacting limit. The bonding-antibonding peaks of the $d_{x^2-y^2}$ orbitals split up by $2h_{x^2-y^2}\approx 1.5$~eV, while the $d_{yz}$ peaks at $-0.03$ and $0.51$~eV are split by $2h_{yz} \approx 0.54$~eV and the splitting of the $d_{xz}$ peaks is very small ($2h_{xz} \approx 0.07$~eV). The $d_{xz}$ peaks and the bonding state of the $d_{yz}$ orbital are close to the Fermi energy. As the interactions are turned on ($a>0$), the $d_{xz}$ and $d_{yz}$ orbitals are pushed up, see panels (b)-(f), and satellite peaks are created around $-2$~eV ($d_{x^2-y^2}$ orbital) in the occupied part of the spectrum, and also about $2$~eV above the unoccupied antibonding peaks ($d_{x^2-y^2}$, $d_{xz}$, $d_{yz}$ orbitals). Combing the information from Figs.~\ref{fig:spectra_ED_hopping} and \ref{fig:spectra_ED_interaction}, we conclude that both interaction effects and bonding-antibonding splittings play important roles in shaping the electronic structure of the realistic system, while the gap size is determined by the inter-orbital same-spin Hubbard interaction $U-3J$. 

Among all states in the Hilbert space of a three-orbital Kanamori-Hubbard dimer, we are interested in the sector with $n=2$ electrons. The ground state energy and wave function of a Hubbard dimer with interaction $U$ and hopping $h$ are $\frac{U-\sqrt{U^2+16h^2}}{2}$ and $|\psi_{\text{GS}}\rangle = \frac{\lambda}{\sqrt{1+\lambda^2}}|s\rangle+\frac{1}{\sqrt{1+\lambda^2}}|d \rangle$, with $\lambda = \frac{4h}{-U+\sqrt{U^2+16h^2}}$ and $|s\rangle$ and $|d\rangle$ defined as in the main text. For our parameter set $U=2.2$ and $h=h_{x^2-y^2} = 0.75$, $|\psi_{\text{GS}}\rangle = 0.89|s\rangle+0.45|d \rangle$. In the realistic system with crystal field splittings, the energy difference between the ground state and the  {\it orbital-mixed doublon state} with energy $U-J+\Delta E_{\text{loc}}$ is roughly $U-J+\Delta E_\text{loc}-\frac{U-\sqrt{U^2+16h^2}}{2} \approx 2.8$, where $\Delta E_{\text{loc}}=0.08$~eV is the energy level splitting between the $d_{x^2-y^2}$ and $d_{xz}$ orbital.

\clearpage

\section{Nonequilibrium spectra}

\begin{figure}[ht]
	\centering
	\includegraphics[width=1\linewidth]{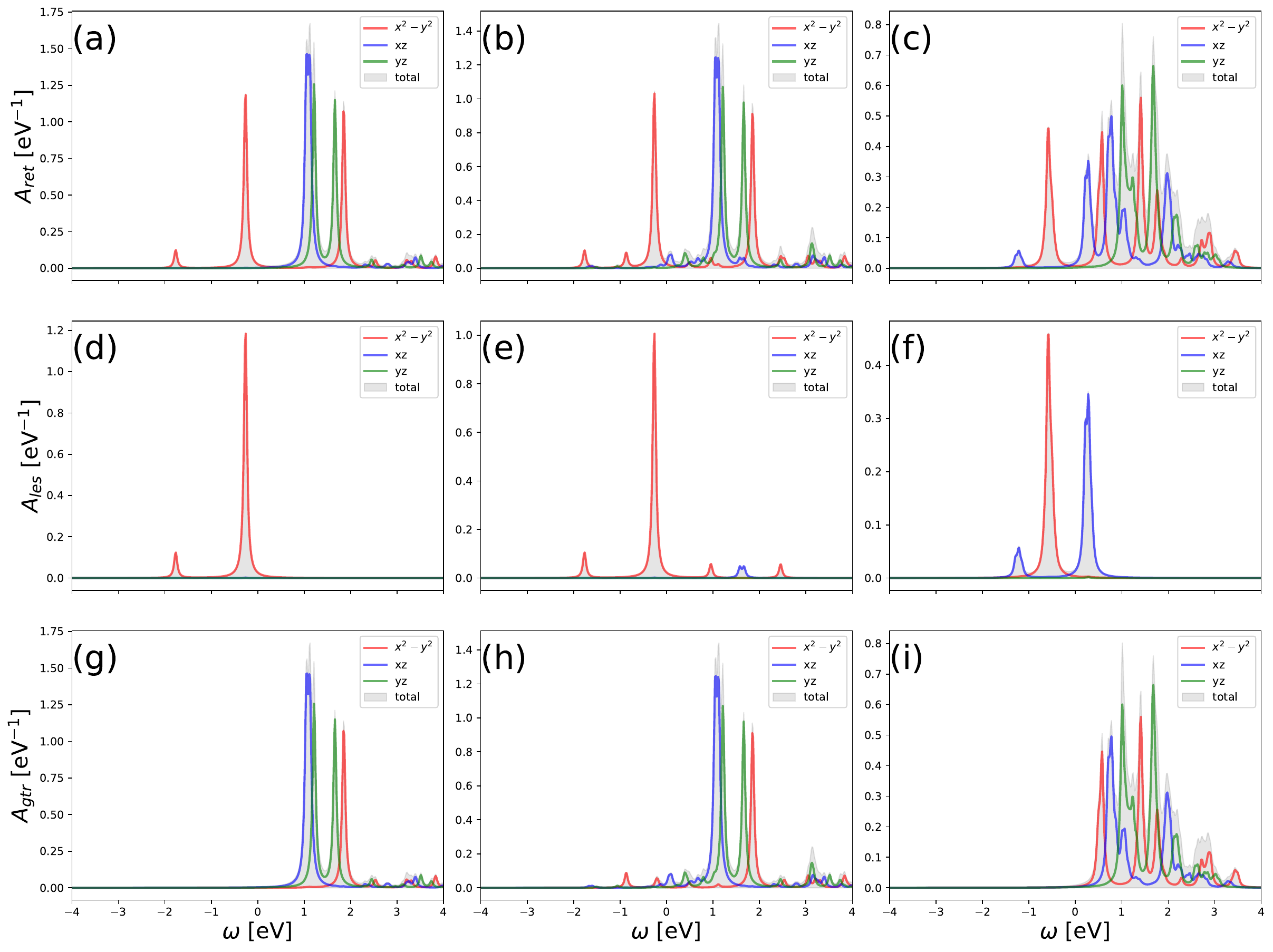}
	\caption{ED spectra of the ground state (left), the excited state after the pulse (middle) and the orbital-mixed singly occupied states (right). From top to bottom: retarded, lesser and greater spectra. A pulse with frequency $\omega_0=2.8$~eV is used.}
	\label{fig:ED_spectra}
\end{figure}

In Fig.~\ref{fig:ED_spectra}, we present the ED spectra for the ground state, the excited state after the pulse, and the {\it orbital-mixed singly occupied state}. As discussed before, in Fig.~\ref{fig:ED_spectra}(a,d,g), the ground state has a gap between the $d_{xz}$ and $d_{x^2-y^2}$ orbital, which is determined by the inter-orbital interaction. The peaks below the Fermi energy are associated with the creation of single-electron $d_{x^2-y^2}$ bonding states (main peak) and anti-bonding states (satellite generated by the interactions). In panels~(b,e,h), we present the spectra for the excited state after the laser excitation. This state is a superposition of the ground state ($\sim87\%$ weight) and the {\it orbital-mixed doublon state} ($\sim12\%$ weight). In panel (e), we observe two peaks in the $d_{x^2-y^2}$ occupation at 2.45~eV and 0.95~eV, and a peak in the $d_{xz}$ occupation around 1.6~eV, which correspond to transitions from the doublon state to a singly occupied state. As discussed in the main text, these peaks are also captured by the cDMFT simulation, but only as transient features when $t\approx t_0$, indicating a very short lifetime of the orbital-mixed doublon state. In panels~(c,f,i), we present the ED spectra of the {\it orbital-mixed singly occupied states}. The relevant observation here are the in-gap peaks contributed by both the $d_{xz}$ and $d_{x^2-y^2}$ orbitals.
	
\end{document}